\documentclass[%
 aip,
 amsmath,amssymb,
reprint,
floatfix]{revtex4-1}

\usepackage{graphicx}
\usepackage{dcolumn}
\usepackage{bm}

\usepackage[utf8]{inputenc}
\usepackage[T1]{fontenc}
\usepackage{mathptmx}
\usepackage{etoolbox}

\usepackage{siunitx}
\DeclareMathOperator{\Tr}{Tr}
\newcommand\Tstrut{\rule{0pt}{2.6ex}}         
\newcommand\Bstrut{\rule[-0.9ex]{0pt}{0pt}}
\usepackage{hyperref}
\usepackage{xcolor}
\hypersetup{
    colorlinks,
    linkcolor={red!50!black},
    citecolor={blue!50!black},
    urlcolor={blue!80!black}
}
\usepackage{ragged2e}
\usepackage[caption=false]{subfig}

\usepackage{float}
\usepackage[normalem]{ulem}

\makeatletter
\def\@email#1#2{%
 \endgroup
 \patchcmd{\titleblock@produce}
  {\frontmatter@RRAPformat}
  {\frontmatter@RRAPformat{\produce@RRAP{*#1\href{mailto:#2}{#2}}}\frontmatter@RRAPformat}
  {}{}
}%
\makeatother
\begin{document}

\preprint{}

\title{Internal Conversion Rates from the Extended Thawed Gaussian Approximation: Theory and Validation}

\author{Michael Wenzel}
 \affiliation{Institut für Physikalische und Theoretische Chemie, Universität Würzburg, Emil-Fischer Str. 42, 97074 Würzburg, Germany
}%

\author{Roland Mitric$^*$}%
\email{roland.mitric@uni-wuerzburg.de}
\affiliation{Institut für Physikalische und Theoretische Chemie, Universität Würzburg, Emil-Fischer Str. 42, 97074 Würzburg, Germany
}%

\date{\today}

\begin{abstract}
The theoretical prediction of the rates of nonradiative processes in molecules is fundamental to assess their emissive properties. In this context, global harmonic models have been widely used to simulate vibronic spectra as well as internal conversion rates and to predict photoluminescence quantum yields. However, these simplified models suffer from the limitations that are inherent to the harmonic approximation and can have a severe effect on the calculated internal conversion rates. Therefore, the development of more accurate semiclassical methods is highly desirable. 
Here, we introduce a procedure for the calculation of  nonradiative rates in the framework of the time-dependent semi-classical Extended Thawed Gaussian Approximation (ETGA).  
 We systematically investigate the performance of the ETGA method by comparing it to the the adiabatic and vertical  harmonic method, which belong to the class of widely used global harmonic models. Its performance is tested in potentials that cannot be treated adequately by global harmonic models, beginning with Morse potentials of varying anharmonicity followed by a double well potential. The calculated radiative and nonradiative internal conversion rates are compared to reference values based on exact quantum dynamics. We find that the ETGA has the capability to predict internal conversion rates in anharmonic systems with an appreciable energy gap, whereas the global harmonic models prove to be insufficient.
\end{abstract}

\maketitle


\section{\label{sec:level1}Introduction}

The efficiency of light-emitting molecular materials depends on the interplay between the radiative and nonradiative relaxation processes such as internal conversion (IC) and intersystem crossing (ISC) that compete with the emission of photons and can be detrimental to technological applications.  For example, the function of light-emitting diodes (LEDs) in displays as well as fluorescence markers used as sensors in chemical or biological systems depend on efficient luminescence.\cite{Frangioni2003,Kim2003,Kuang2003,Peng2011,Shuai2016,Zhang2018,Hudson2021,Song2022}   
Theoretical methods can directly contribute to technological advancement by virtual pre-screening of potential molecules, according to their properties like emission wavelength and fluorescence quantum yield. This requires efficient methods for the simulation of absorption and emission spectra as well as for predicting the rates of radiative and nonradiative relaxation processes.
In this context, the prediction of the internal conversion rates is particularly challenging since the IC process can take place on an ultrafast timescale if the molecular system has accessible conical intersections connecting electronic states. This often leads to complex dynamics involving an intricate interplay of electronic and nuclear degrees of freedom. A proper treatment of such nonadiabatic dynamics requires methods that go beyond the Born-Oppenheimer approximation. A fully quantum mechanical treatment is possible using  linear vibronic coupling models combined with the multilayer multiconfigurational time-dependent Hartree method (ML-MCTDH).\cite{Wang2003,Manthe2008} While computationally demanding, this approach can handle strongly coupled electronic states and reveal their short time dynamics.\cite{Liu2018} A second promising approach is the time-dependent density matrix renormalization group method (TD-DMRG).\cite{Green2021,Wang2021,Ren2021} 

An attractive alternative to these exact approaches are  mixed quantum-classical methods using classical trajectories that are allowed to hop between adiabatic electronic states, feasible even in  systems that are too costly for a full quantum mechanical treatment.\cite{Tully1990,Mitric2009,Barbatti2011,Roehr2013,Hoche2017,Malhado2014,Cui2014,Xie2019} 

The nonradiative decay of excited states can be slowed down by energetic barriers. In these cases the access to the conical intersection is the rate-determining step and can be treated using Kramers reaction rate theory.\cite{Blacker2013,Hoche2019,Kramers1940,Kohn2019,Lin2020} It is also possible that systems do not relax through a conical intersection but simply due to small, almost constant kinetic couplings. In this case, the nonradiative relaxation can be modelled within the time-dependent perturbation theory e.g employing Fermi's golden rule.
Since systems in this limit are expected to show non-negligible fluorescence that is not suppressed by an ultrafast nonradiative decay, it would be desirable to have a theoretical approach for predicting the fluorescence quantum yields that relies on the perturbation theory but is more accurate than the widely used global harmonic models.\cite{Hazra2003,Hazra2004,Dierksen2005,Santoro2008,Barone2009,Bloino2010,Ferrer2012,Sorour2022} The expressions for the internal conversion rate within the perturbation theory framework are based on Fermi's golden rule \cite{Dirac1927,Fermi1950}  and existing methods differ in their approach and the approximations made to evaluate it. The prominent examples are the analytic energy gap law and its more recent  variants that are applicable even to large molecular systems.\cite{Englman1970,Plotnikov1979,Valiev,Valiev2018,Valiev2021,Hoche2019,Erker2022}

Another approach is to evaluate the sum-over-states rate expressions within a global harmonic approximation for the involved electronic potential energy surfaces.\cite{Peng2007,Santoro2008,Niu2010} All required matrix elements are then available as analytic expressions but the quickly growing number of terms due to the high amount of possible vibronic transitions can make the summation cumbersome. A time-domain approach eliminates the need to evaluate the summation of matrix elements and replaces the problem with the task of propagating a nuclear wave packet to obtain an autocorrelation function. Analytic expressions for this autocorrelation function are also available within the global harmonic approximation.\cite{Hayashi1998,Peng2013,Banerjee2016,Miyazaki2022}

However, the internal conversion usually occurs from a low lying vibrational state of the initial potential energy surface to a highly excited vibrational state of the final electronic potential.  The prediction of the internal conversion rate may thus require an accurate treatment of higher lying vibrational states, which is not possible without an adequate handling of the anharmonicities of the underlying potential, which are completely absent in global harmonic models.\cite{Ianconescu2011}
An attractive possibility to tackle this problem is provided by the  various semi-classical methods for wave packet propagation.\cite{lasser_lubich_2020, Heller1975,Heller1976,Heller1981,Heller1981a,Hagedorn1980,Hagedorn1985,Hagedorn1998,Herman_1984,Kluk1986,Walton1996} The basic idea is to express the initial state as a sum of Gaussian wave packets that are guided by classical trajectories, acquiring a phase factor according to the corresponding classical action. Differences exist in the details of the parametrization of the used Gaussian functions. They can be allowed to vary in width or remain frozen during the propagation.\cite{Heller1981a,Kluk1986} A major benefit is that only local information of the potential is required. Thus,  it is not necessary to precompute the potential energy surface as is the case for quantum dynamics simulations. It was shown that this approach is applicable to the calculation of internal conversion rates in anharmonic potentials, using a semi-classical initial value representation method based on swarms of Gaussian wave packets.\cite{Ianconescu2011,Ianconescu2013} This approach can be used to model the dynamics of initial wave packets of arbitrary shape but may require many trajectories to converge.

In this work we introduce a method for the calculation of the internal conversion rates in the frame of the Extended Thawed Gaussian Ansatz (ETGA) that was recently used in a series of publications to predict vibronically resolved spectra.\cite{Patoz2018,BEGUSIC2018_3TGA,Begusic2019,Prlj2020,Begusic2020,Begusic2022}
In this approach, the anharmonicity of a potential is automatically imprinted into the guiding trajectory and impacts the wave packet dynamics, which enables the method to treat cases where global harmonic models reach their limit.
We explore the  viability of the ETGA as tool for the prediction of radiative and nonradiative quantum yields in one-dimensional model systems that can be treated by exact numerical calculations. Furthermore, we compare systematically the performance of the ETGA with the global harmonic models. We begin by exploring the capabilities and limits of the ETGA model using Morse potentials with varying degrees of anharmonicity. This is followed by an application of the method to a double well potential with varying starting positions of the wave packet. We wish to emphasize that although one dimensional models are certainly not representative for complex polyatomic molecules the differences between  approximations for calculation of rates are most clearly seen in one dimension, where exact results can be easily obtained. 


\section{Theory}

The ETGA method is derived by making the following ansatz for the time-dependent nuclear wave packet,\cite{Lee1982}

\begin{equation}
    \phi_t(q) = \exp \left(i \left[\frac{1}{2}(q-q_t)^TA_t(q-q_t)+p^T_t(q-q_t) + \gamma_t \right] \right)\,, \label{eq:tga_definition}
\end{equation}

where all quantities are given in mass-weighted normal coordinates and Hartree atomic units. The time dependence is incorporated in the parameters \(A_t,q_t, p_t\) and \(\gamma_t\), which satisfy the following differential equations 

\begin{align}
    \dot q_t &= p_t \quad \dot p_t = -V'_t \label{eq:eom_qp_t} \\
    \dot A_t &= -A_t A_t - V''_t\label{eq:eom_A_t} \\
    \dot \gamma_t &= i\frac{1}{2}\Tr(A_t) + L_t\label{eq:eom_gamma_t},
\end{align}
that can be derived by inserting the wave packet ansatz into the time-dependent Schrödinger equation with a modified potential.\cite{Heller1975} The original potential is replaced by a time-dependent local harmonic approximation of the form
\begin{equation}
V_t(q) = V(q_t) + V'_t (q-q_t) + \frac{1}{2}(q-q_t)^T V''_t(q-q_t)\,\label{eq:lha_potential}.
\end{equation}
Solving differential equations (\ref{eq:eom_qp_t}--\ref{eq:eom_gamma_t}) is equivalent to the propagation of a  Gaussian wave packet using an effective time-dependent Hamiltonian\cite{Heller1976}
\begin{equation}
H_t = -\frac{1}{2}\nabla_q^2 + V_t(q)
\end{equation}
such that
\begin{equation}
U\phi_0 = \phi_t,
\end{equation}
with \(U = \mathcal{T} \exp(-i \int_0^t H_{t'} dt') \), introducing the time ordering operator \(\mathcal T\). Since the parameters of the local harmonic potential are all time-dependent, such evolving potential can in principle accurately approximate any globally anharmonic potential energy surface. We will assume in the following that the initial state has a Gaussian shape, which is a good approximation for the vibrational groundstate of a system in equilibrium. A single Gaussian function with time-dependent parameters will then suffice, requiring only a single trajectory to be run. The parameters \(q_0,p_0, A_0\) and \(\gamma_0\) are determined by comparison with the initial vibrational ground state wave function.

The mass-weighted normal coordinate \(q_t \) and normal momentum \(p_t \) follow the classical equations of motion (\ref{eq:eom_qp_t}) and are obtained  by running a classical trajectory on the original fully anharmonic adiabatic potential energy surface \( V(q) \). This requires evaluations of the gradient \((V'_t)_i= \frac{\partial V(q_t)}{\partial q_i} \,,\) while the integration of equation (\ref{eq:eom_A_t}) for \( A_t \) necessitates the calculation of the Hessian matrix \( (V''_t)_{i,j} = \frac{\partial^2 V(q_t)}{\partial q_i \partial q_j}\,.\) Parameter \( \gamma_t \) contains a classical contribution based on the Lagrangian \( L_t = \frac{1}{2}p_t\cdot p_t - V_t \) evaluated along the trajectory \( q_t \) and ensures that the wave packet remains normalized.

The internal conversion rate will be calculated with the assumption that our system is initially in the vibrational ground state \( \phi_{i0} \) of an adiabatic electronic excited state \( |i\rangle  \), which is assumed to be harmonic, close to its equilibrium configuration. The kinetic energy operator for nuclei, \( \hat T \), is treated like a constant perturbation, facilitating a radiationless de-excitation to excited vibrational states of the adiabatic electronic ground state. We consider all transitions from a single initial vibronic state \(\phi_{i0}\) to the set of vibronic eigenstates \(\{\phi_{fj}\} \) of the final potential energy surface, where the first and second indices indicate the electronic and vibrational state, respectively.

The transition rate according to Fermi's golden rule\cite{Ianconescu2011,Ianconescu2013,Humeniuk2020} is given by 

\begin{equation}
k_{\mathrm{IC}} = 2\pi \sum_j |\langle \phi_{fj}|T_{fi}|\phi_{i0}\rangle|^2\delta(\omega_{i0}-\omega_{fj}).\label{eq:fgr_ti}\\   
\end{equation}
We switch to the equivalent time-dependent picture\cite{Tannor2006} by introducing the Fourier integral representation of the Dirac delta distribution \(\delta(\omega)=\frac{1}{2\pi}\int^\infty_{-\infty} dt \exp(i\omega t) \) into Eq. (\ref{eq:fgr_ti}),
\begin{equation}
k_{\mathrm{IC}} =\int^\infty_{-\infty} dt \exp(i\omega_{i0}t) \langle  \phi_{i0}|T^\dagger_{fi} \sum_j | \phi_{fj}\rangle  \exp(-i\omega_{fj}t) \langle \phi_{fj}|T_{fi}|\phi_{i0}\rangle. 
\end{equation}
 Noticing that \(\exp(i\omega_{i0} t)\langle \phi_{i0}| =\langle \phi_{i0}|U_i^\dagger \) and using the the spectral representation of the propagator associated with the final electronic state, \(U_f = \sum_j \exp(-i\omega_{fj}t) |\phi_{fj}\rangle \langle \phi_{fj} |\), leads to the following expression for the internal conversion rate
\begin{equation}
 k_{\mathrm{IC}} = \int^\infty_{-\infty} dt  \langle \phi_{i0}|U^\dagger_iT^\dagger_{fi} U_f T_{fi}|\phi_{i0}\rangle.\label{eq:kic}
\end{equation}
 
  The matrix element \(T_{fi} \) of the nuclear kinetic energy operator \(\hat T = -\frac{1}{2}\nabla^2_q \) with respect to an orthogonal adiabatic electronic basis in mass-weighted normal coordinates is given by
\begin{align}
 T_{fi} &= \langle f | \hat T | i  \rangle_r = -\langle f |\nabla _q i\rangle_r^T \nabla_q - \frac{1}{2}\langle f |\nabla^2 _q i\rangle_r 
\end{align} 
We assume that the second-order nonadiabatic coupling \(\langle f |\nabla^2 _q i\rangle_r\) can be neglected and use a constant value for the first-order nonadiabatic coupling  \(\tau_{fi} = \langle f |\nabla _q i\rangle_r \). This simplifies the kinetic coupling to the following expression
\begin{equation}
 T_{fi} =- \tau_{fi}^T\nabla_q.  
\end{equation}

We will now focus on the core of the problem, that is the evaluation of the term
\begin{equation}
    U_f |T_{fi}\phi_{i0}\rangle 
\end{equation}
which, besides some constant factors, is identical to the task of propagating the gradient of the initial thawed Gaussian.  The gradient of Eq. (\ref{eq:tga_definition}) is readily evaluated and leads to
\begin{equation}
    \begin{aligned}
        \nabla_q \phi_{i0} &= (iA_0(q-q_0) + ip_0)\phi_{i0}.\label{eq:q_dependent_etga} \\
    \end{aligned}
\end{equation}
The time evolution operator \(U_f\) does not commute with \(q\). But it can be shown that it commutes with the initial momentum derivative operator \(\nabla_{p_0}\) within the local harmonic approximation for the potential.\cite{BEGUSIC2018_3TGA} We can make use of this fact to rewrite  Eq. (\ref{eq:q_dependent_etga}) using the gradient with respect to the initial momentum
\begin{equation}
        \nabla_q \phi_{i0} = (A_0\nabla_{p_0} + ip_0)\phi_{i0}\label{eq:p0_dependent_etga}. \\
\end{equation}
The propagator can now act directly on the initial wave packet, 
\begin{equation}
    \begin{aligned}
        U_f(t) \nabla_q \phi_{i0}(t_0) &= U_f(t) (A_0 \nabla_{p_0} + ip_0)\phi_{i0}(t_0)\\
        &= (A_0 \nabla_{p_0} + ip_0)U_f(t)\phi_{i0}(t_0) \\
        &= (A_0 \nabla_{p_0} + ip_0)\phi_{i0}(t).
    \end{aligned}
\end{equation}

 The time evolution of the wave packet is thus obtained by integration of the parameters that have been defined in Eqs. (\ref{eq:eom_qp_t})--(\ref{eq:eom_gamma_t}). The only additional step required is the evaluation of the gradient with respect to the initial momentum, which, as derived in Appendix D of reference \onlinecite{BEGUSIC2018_3TGA}, yields the following result
\begin{align}
    \nabla_{p_0}\phi_{i0}(t) &= i(M_{t,pp}^T -M_{t,qp}^TA_t)(q-q_t)\phi_{i0}(t),
\end{align}
where we have introduced elements of the monodromy matrix \(M_t\) defined by
\begin{equation}
M_t= \left( \begin{matrix}  M_{qq} & M_{qp} \\ M_{pq} & M_{pp} \end{matrix} \right) =   \left( \begin{matrix}  \frac{\partial q_t}{\partial q_0} & \frac{\partial q_t}{\partial p_0 } \\ \frac{\partial p_t}{\partial q_0} & \frac{\partial p_t}{\partial p_0} \end{matrix} \right) \label{eq:monodromy_definition}, \\
\end{equation}
with \( \left( \frac{\partial q_t}{\partial q_0} \right)_{i,j}  = \frac{\partial q_i(t,q_0,p_0)}{\partial q_{0,j}}\,. \) The matrix is obtained by integration of the following equation\cite{Huber1988,Zhuang2012}:
\begin{equation}
\dot M_t =\left( \begin{matrix}  0 & 1 \\ -V''_t & 0 \end{matrix} \right) M_t.
\end{equation}

Assuming that internal conversion is the only competing decay channel allows the calculation of the quantum yield of spontaneous emission \(\Phi\) as

\begin{equation}
 \Phi = \frac{k_\textrm{SE}}{k_\textrm{SE}+k_\textrm{IC}}  \label{eq:quantum_yield},
\end{equation}

where the spontaneous emission rate \(k_\textrm{SE}\) is obtained by integration of the emission spectrum \(\sigma_\mathrm{SE} \),\cite{Niu2010,Begusic2019}

\begin{equation}
k_\textrm{SE} = \int_0^\infty d\omega \ \sigma_\mathrm{SE}
\end{equation}

with 

\begin{equation}
\begin{aligned}
\sigma_\textrm{SE}(\omega) &= \frac{\omega^3 }{6\pi^2\hbar \varepsilon_0 c^3} \int^\infty_{-\infty} dt  \langle \phi_{i0} | U^\dagger_i \mu^\dagger_{fi}U_f \mu_{fi}|\phi_{i0}\rangle \exp(-i\omega t). \\[1.0em]
\end{aligned}
\end{equation}

The transition dipole moment is defined as \({\mu_{fi}(q) \equiv \mu_0 + \mu'(q-q_0)}\), including the Franck-Condon and Herzberg-Teller term.\cite{Franck1926,Condon1928,Herzberg1933}

Furthermore we will use the internal conversion correlation function from Eq. (\ref{eq:kic})
\begin{equation}
    C(t) =  \langle \phi_{i0} | U^\dagger_i T^\dagger_{fi}U_f T_{fi}|\phi_{i0}\rangle
\end{equation}
to define \(k_\mathrm{IC}(\omega)\), the internal conversion spectrum
\begin{equation}
    k_\mathrm{IC}(\omega) =  \int^\infty_{-\infty} dt \ C(t) \exp(i\omega t)\label{eq:ic_spectrum}.
\end{equation}
The angular frequency \( \omega \) can be interpreted as the energy gap between the initial and final state and the internal conversion rate is contained in this spectrum at the point of energy conservation, \( \hbar\omega=0 \ \mathrm{eV}\). Plots of the internal conversion rate as a spectrum show how the rate would change with the energy gap of the initial and final potential and peaks appear when \(\hbar\omega \) matches the energy difference between states of the final and initial potential. Peaks at negative frequencies indicate states that are lower in energy than the initial state and those at positive energy correspond to higher lying states. This spectrum unveils all the information that is stored in the time-dependent correlation function and provides more information than the single valued rate by itself. It also constitutes a better measure to gauge and compare the quality of different methods.

\section{Morse Potential}

\subsection{Internal Conversion}

The ETGA is exact for harmonic potentials since the local harmonic expansion is identical to the true potential in that case. A potential with anharmonicity is thus required to test the performance of the ETGA, with the Morse potential being ideally suited as its anharmonicity can be varied continuously.\cite{Morse1929,Humeniuk2020} The results of the ETGA are compared to the results obtained using a split-operator propagation scheme\cite{Feit1982,Feit1983,Feit1984,Kosloff1988} (SOP) with the true potential as well as to the adiabatic and vertical harmonic approximations\cite{Hazra2003,Hazra2004,Ferrer2012} (AH and VH) in three Morse potentials with increasing anharmonicity. The initial potential is always the same and assumed to be harmonic and written in the form 
\begin{equation}
    V_i(q)= \frac{1}{2}\omega^2(q-q_{i,0})^2 + V_{i,0},
\end{equation}
the final potential is defined as 
\begin{equation}
    V_f(q) = D\left(1 - e^{-\frac{\omega}{\sqrt{2D}}(q-q_{f,0})}\right)^2 + V_{f,0}.
\end{equation}
The anharmonicity of the Morse potential\cite{Humeniuk2020} is defined as 
\begin{equation}
\chi = \frac{\omega}{4D}\label{eq:anharmonicity}.
\end{equation}

\begin{figure}
  \centering
  \includegraphics[width=.50\textwidth, trim=0 0.25cm 0 0.5cm, clip]{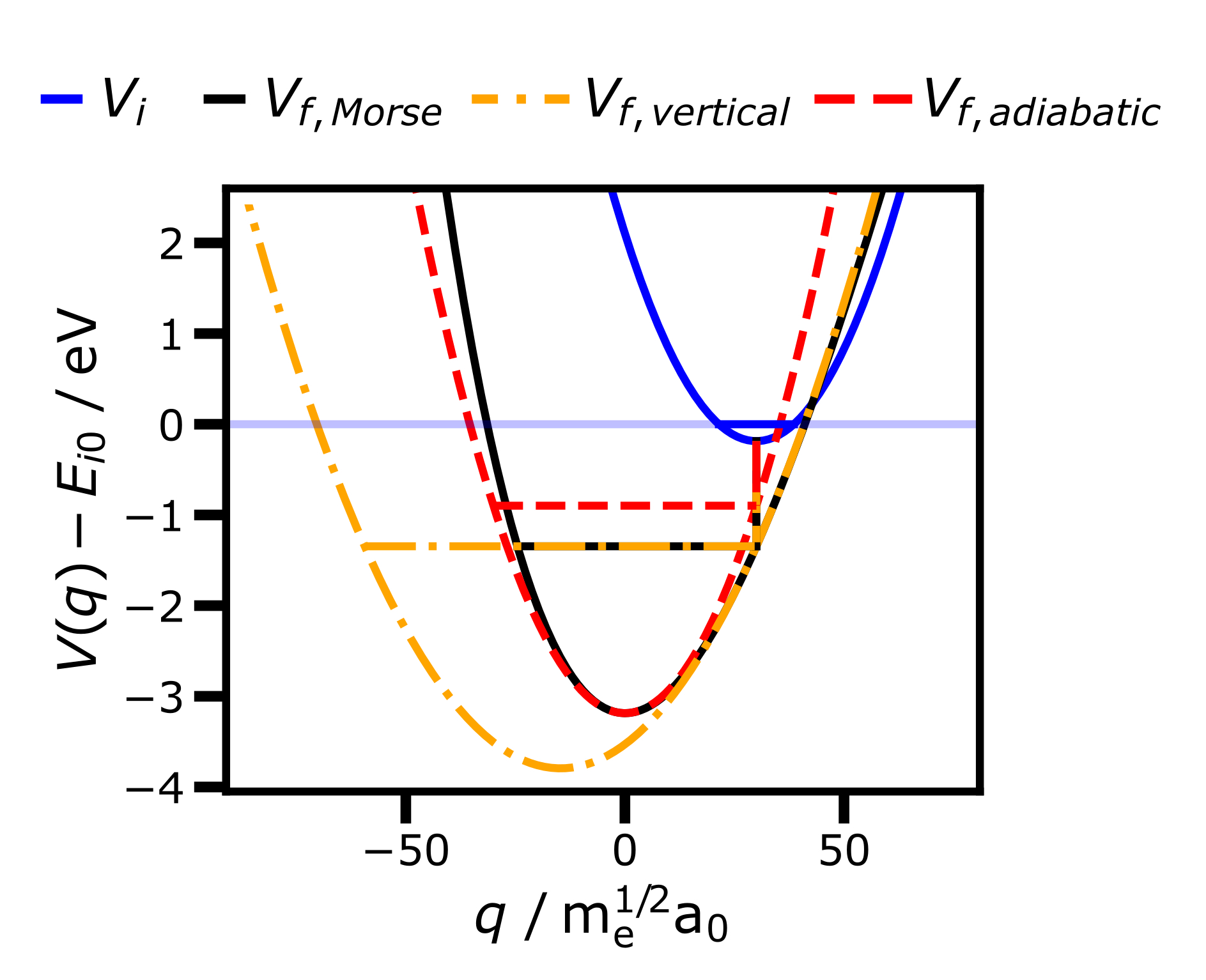}
\vspace{-20pt}
  \caption{Initial (blue) and final (black) potential, together with the vertical (orange) and adiabatic (red) harmonic approximation to the final potential. The final potential is a Morse potential with anharmonicity \(\chi=0.002\). The energy axis is shifted by \(E_{i0}\), the eigenvalue of the initial vibrational eigenstate of harmonic potential \(V_i\). Horizontal lines in the same color and linestyle as the potentials indicate the starting and turning points of classical trajectories that start at the equilibrium position of \(V_i\).}
  \label{fig:low_anharmonicity_potentials}
\end{figure}

The fundamental frequency is the same for the initial and final potential and is kept constant at \(\omega = 3000 \textrm{ cm}^{-1}\).
The minima of the potentials are \(30 \ \textrm{m}_\textrm{e}^{1/2}\textrm{a}_\textrm{0}\) mass weighted units apart.
The first order kinetic coupling parameter is in all cases \({\tau_{fi} = 1.0 \ (\textrm{m}_\textrm{e}^{1/2}\textrm{a}_\textrm{0})^{-1}  }\). The adiabatic energy difference of the potentials is set constant at 3.0 eV. We begin with an anharmonicity of \( \chi=0.002 \) corresponding to a well depth \({D = 46.5 \textrm{ eV}}\). The initial harmonic potential and the Morse potential are shown in Figure \ref{fig:low_anharmonicity_potentials}, together with the vertical and adiabatic harmonic approximations to the final potential. The initial potentials in our models are always assumed to be a harmonic approximation to an adiabatic electronic potential and as such may cross the final potential while working in an adiabatic basis with derivative couplings. The turning points of a classical trajectory starting at the equilibrium position of the initial potential are indicated by horizontal lines. 

\begin{figure*}
  \centering
  \includegraphics[width=.95\linewidth]{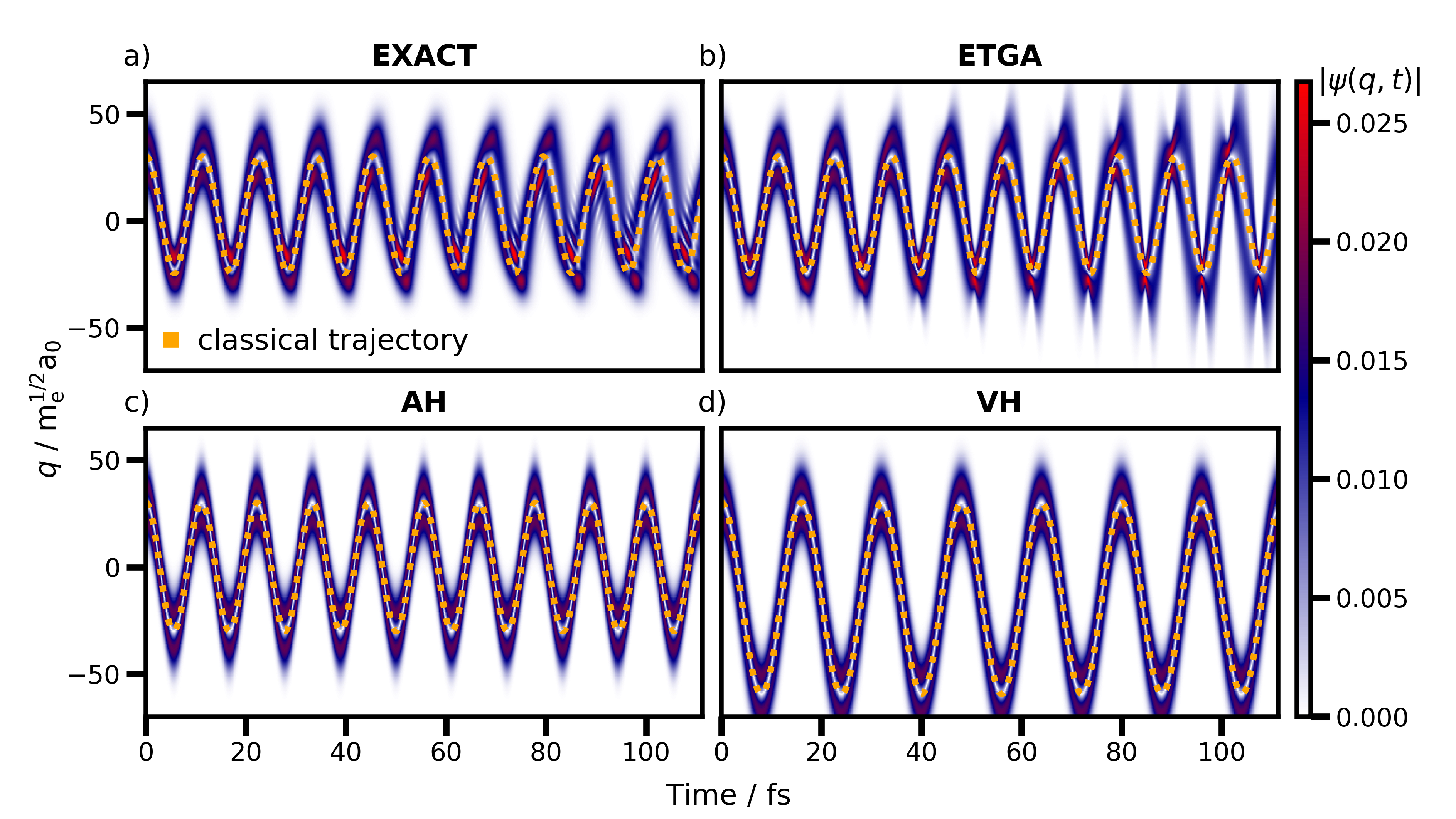}
  \caption{Magnitude of the wavepacket \(\psi(q,t) =  \langle q| U_f T_{fi}|\phi_{i0}\rangle \) obtained by numerical propagation using a split-operator scheme (a), the extended thawed Gaussian Ansatz (b), the adiabatic harmonic approximation (c) and the vertical harmonic approximation (d). The final potential \(V_f\) is a Morse potential with an anharmonicity of 0.002.}
  \label{fig:wp_morse1}
\end{figure*}

The difference between the vertical and adiabatic approximation is clearly visible even in the case of low anharmonicity  and it is a priori not certain which of the harmonic approximations is better suited for the calculation of nonadiabatic transition rates. 
The VH model is able to capture the short-time dynamics more accurately than the AH model, but at the same time the vertical model should have a larger error regarding wave packet recurrence times, which determine the splitting\cite{Heller1981} of individual vibrational states.
The one dimensional model allows a convenient visualization of the wave packet dynamics as shown in Figure \ref{fig:wp_morse1}. 
The center of the wave packet follows closely the classical trajectory (indicated by orange squares), even when using a full quantum mechanical treatment (Fig. \ref{fig:wp_morse1}a) without approximations as long as the anharmonicity is low. But the delocalisation and spreading of the wave packet clearly increase with each reflection at the turning point of the trajectory.
These effects cannot be captured by a single Gaussian, even if its width is varying with time. The VH wavefunction (Fig. \ref{fig:wp_morse1}d) travels to regions that are energetically forbidden in the true Morse potential,  while the AH (Fig. \ref{fig:wp_morse1}c) and ETGA wavefunction (Fig. \ref{fig:wp_morse1}b) appear quite similar to the propagation based on the original potential.

\begin{figure*}
  \centering
  \includegraphics[width=.95\linewidth]{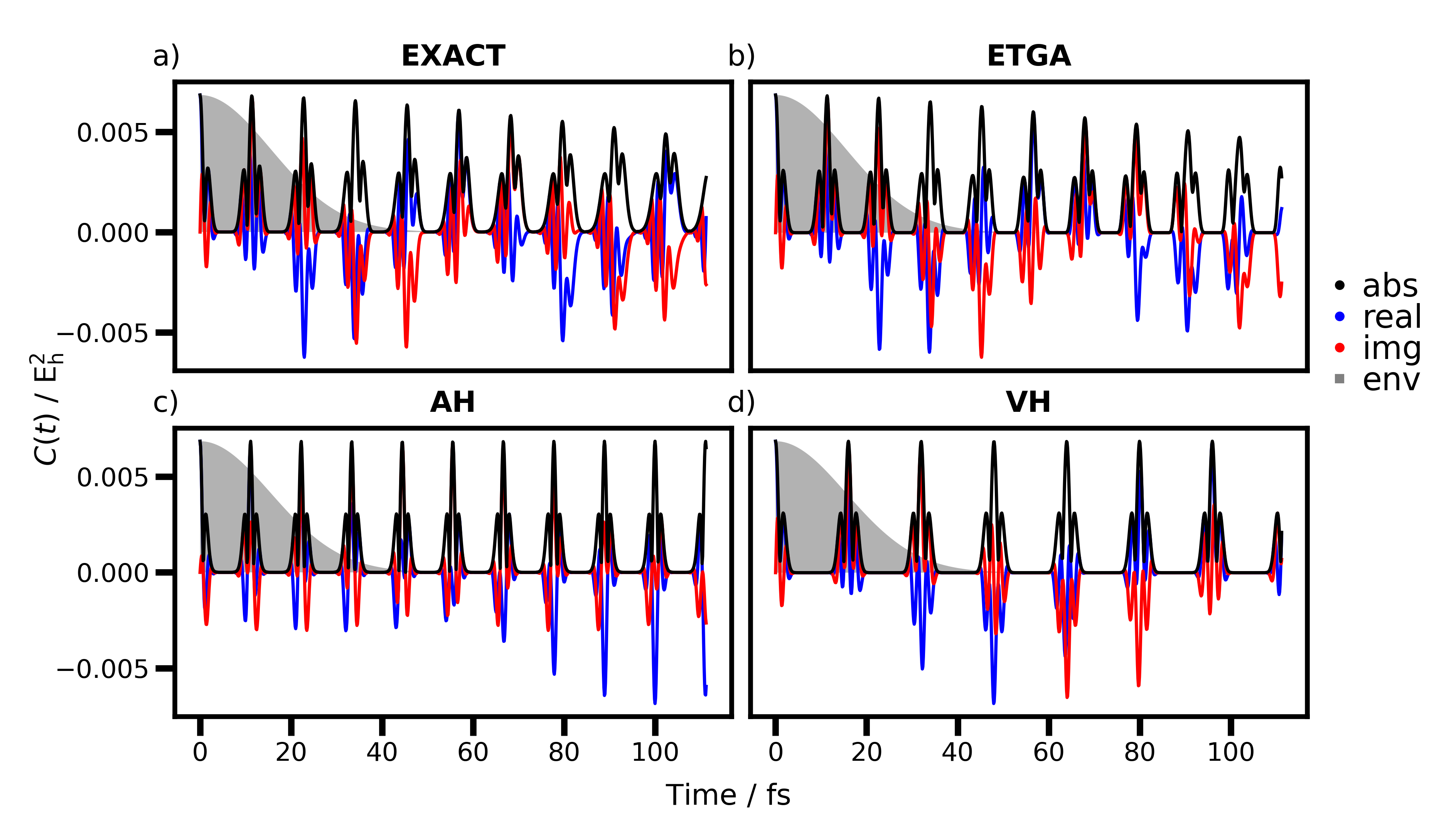}
  \caption{Auto correlation function \(C(t) =  \langle \phi_{i0} | U^\dagger_i T^\dagger_{fi}U_f T_{fi}|\phi_{i0}\rangle\) obtained by numerical propagation using a split-operator scheme (a), the extended thawed Gaussian Ansatz (b), the adiabatic harmonic approximation (c) and the vertical harmonic approximation (d). The final potential \(V_f\) is a Morse potential with an anharmonicity of 0.002. The time-domain Gaussian envelope (env), corresponding to a HWHM of 0.05 eV in the frequency domain, is shown as grey background.}
  \label{fig:cor_morse1}
\end{figure*}

The corresponding autocorrelation functions are provided in Figure \ref{fig:cor_morse1}a--d. Time evolution under the true Morse potential leads to a spreading of the wavepacket and thus a decreasing overlap with the initial wavefunction which expresses itself as slow decay of the correlation function (Fig. \ref{fig:cor_morse1}a). The ETGA can reproduce this behavior (Fig. \ref{fig:cor_morse1}b) unlike the global harmonic models, where the magnitude of the overlap keeps returning to its initial value (Fig. \ref{fig:cor_morse1}c,d). Extraction of the spectral content is provided using a Fourier transform. The correlation function is mirrored and conjugated to extend the signal to negative times. It is then weighted with a Gaussian lineshape function to remove Gibbs artifacts that would occur due to the sudden cutoff of the signal after the finite propagation time of \(T=10\frac{2\pi}{\omega}\approx 111.2 \textrm{ fs}\).

\begin{figure}
  \centering
  \includegraphics[width=.50\textwidth]{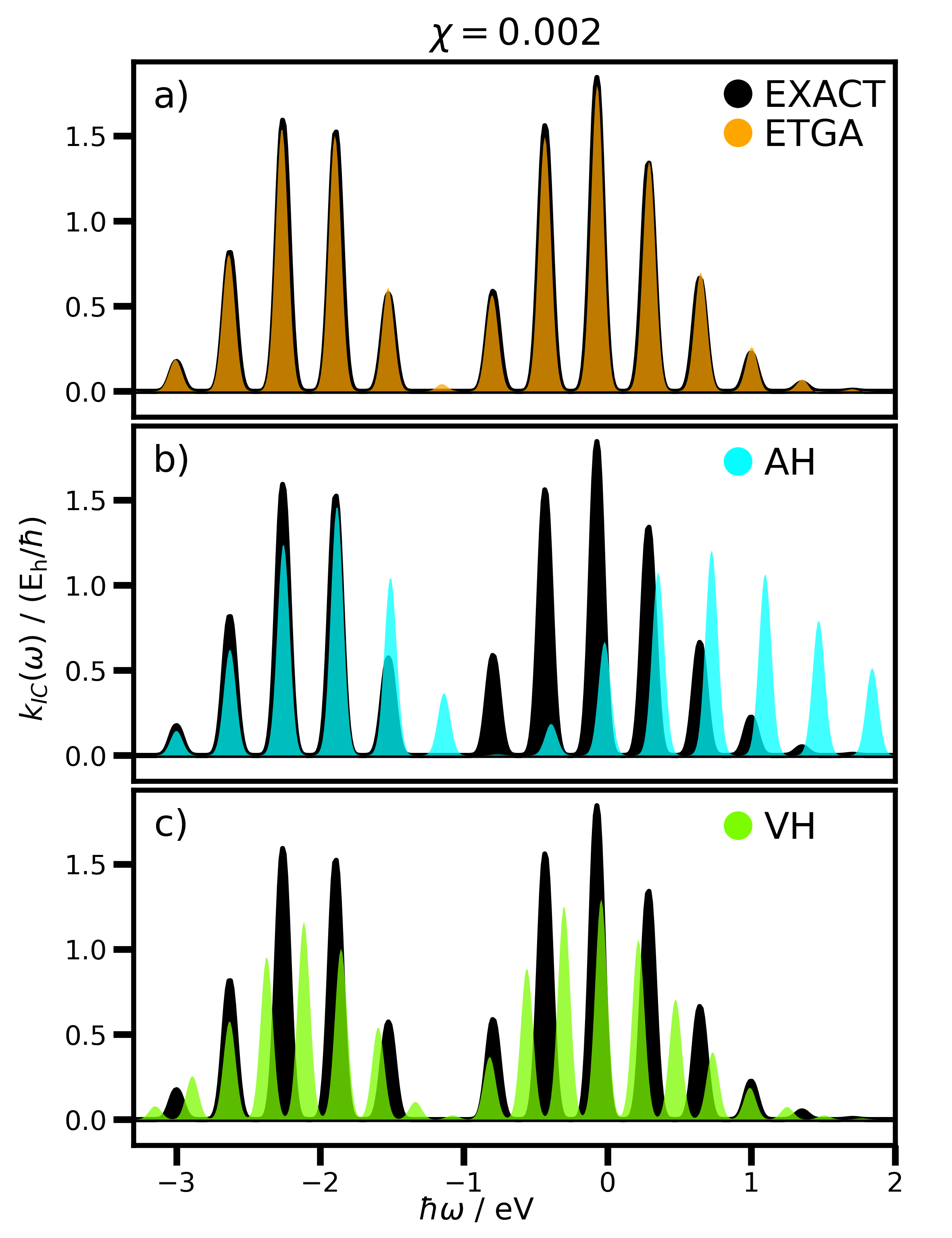}
  \caption{Internal conversion spectrum (Eq. (\ref{eq:ic_spectrum})) based on a Morse potential with anharmonicity of 0.002. The reference spectrum in black shows the results based on the Morse potential without approximations. The spectrum using the ETGA is given in orange in the top panel (a). The middle plot (b) shows the adiabatic harmonic approximation spectrum (cyan) and the bottom (c) the vertical harmonic approximation spectrum (green). The spectra are based on the correlation functions shown in Figure \ref{fig:cor_morse1} and broadened with a Gaussian line shape function with a half width at half maximum (HWHM) of 0.05 eV. The internal conversion rate is the value of the spectrum at \(\hbar\omega=0 \textrm{ eV}\). Spectra are not shifted or scaled to improve agreement with the reference.}
  \label{fig:ic_spectra_low_anharmonicity}
\end{figure}

The spectrum based on the SOP propagation scheme serves as reference in the comparison of the different approximations and is labelled as EXACT in the figures. The top panel of Figure \ref{fig:ic_spectra_low_anharmonicity} shows the reference and the ETGA spectrum. The agreement is very good and the position and height of the peaks are well reproduced over the whole energy range. The adiabatic harmonic model (Fig. \ref{fig:ic_spectra_low_anharmonicity}b) manages to describe the energetically low lying states but fails at higher energies. The vertical model (Fig. \ref{fig:ic_spectra_low_anharmonicity}c) performs better in that region but shows shortcomings in the low energy region. The ETGA is in this case clearly superior to either of the global harmonic approximations.

\begin{figure*}
  \centering
  \includegraphics[width=0.9\textwidth]{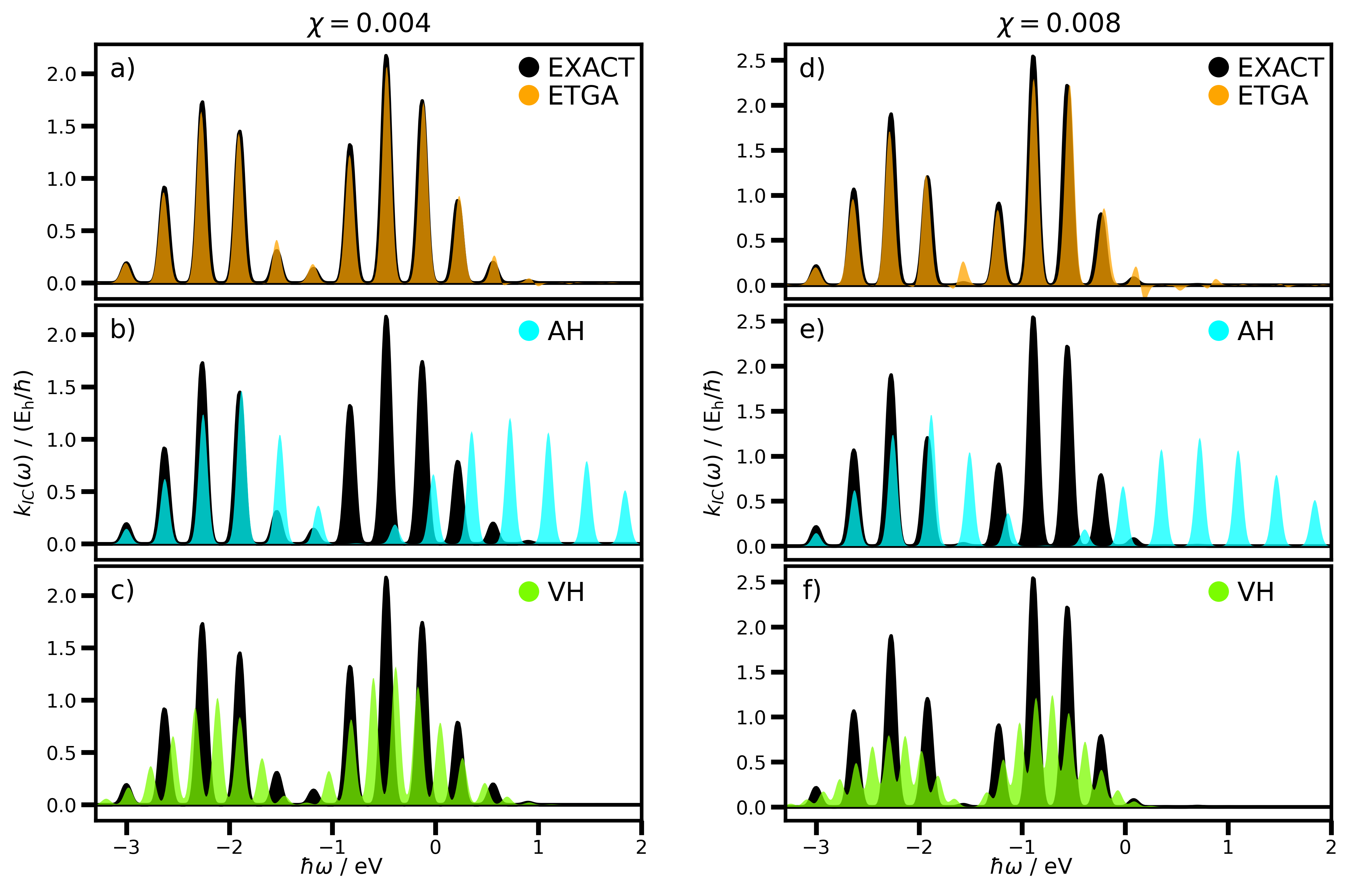}
\caption{Internal conversion spectra based on a Morse potential with anharmonicity of 0.004 (left column, a--c) and 0.008 (right column, d--f). The reference spectrum in black is based on the Morse potential without approximations. The top panels (a,d) show the ETGA spectrum in orange, the adiabatic harmonic model (cyan) is in the middle (b,e) and the vertical harmonic model (green) at the bottom (c,f). All spectra have been broadened using a Gaussian line shape function with a HWHM of 0.05 eV.\label{fig:ic_spectra_morse_med_high}}
\end{figure*}

The reliability of the ETGA was further tested by increasing the anharmonicity of the Morse potential from 0.002 to 0.004 and finally 0.008,  moderate values that ensure that the second derivative of the Morse potential remains positive at the initial geometry, so that the vertical harmonic approximation to the potential remains bound. This corresponds to decreasing values of the well depth parameter \(D\) from 46.5 eV to 23.2 eV and lastly 11.6 eV. 
The spectra for the case with medium anharmonicity of 0.004 are shown in Figure \ref{fig:ic_spectra_morse_med_high}a--c, the case with an anharmonicity of 0.008 in Figure \ref{fig:ic_spectra_morse_med_high}d--f. Increasing the anharmonicity to 0.004 has little effect on the quality of the ETGA spectrum (Fig. \ref{fig:ic_spectra_morse_med_high}a), it is still able to reproduce the exact spectrum remarkably well. But the limitations of the method begin to show in Figure \ref{fig:ic_spectra_morse_med_high}d as the anharmonicity is further increased. Peaks with negative intensity begin to appear as a consequence of the time dependent nature of the approximation, as well as peaks that are nonexistent in the reference. The approximation still matches the reference spectrum to a good degree but it becomes clear that the reliability is diminishing. The adiabatic harmonic approximation is the same for all cases (Fig. \ref{fig:ic_spectra_morse_med_high}b,e) as the second derivative at the minimum stays constant at \(\omega = 3000 \textrm{ cm}^{-1}\) for the Morse potentials. The spacing between consecutive states is constant, leading to an overestimation of the energy gap of excited states that are moving closer together in the true potential. The vertical harmonic approximation (Fig. \ref{fig:ic_spectra_morse_med_high}c,f) has a similar problem. The second derivative evaluated at the starting position of the wave packet corresponds to angular frequencies that decrease as the anharmonicity of the Morse potential is increased. The vertical harmonic potentials take on values for \(\omega\) of 2083, 1742, 1279 \(\textrm{ cm}^{-1}\) for the three anharmonicity parameters 0.002, 0.004, 0.008. This leads to an underestimation of the energy gap for low lying states of the potential and the states are bunching together in the energy range of interest.

The vertical harmonic model can capture the envelope of the vibronic pattern over the whole range of energies but agreement with the exact vibronic pattern cannot be achieved for any of these levels of anharmonicity. In contrast, the adiabatic harmonic model can be trusted for low lying vibrational states and might be suitable if the final and initial states are energetically very close, but its ability to capture the features of the spectrum worsens quickly for higher lying states. Even the general shape begins to deviate notably at higher energies. One can get either the envelope right by using the VH model or the transitions to the lowest vibrational states using the AH model. The extended thawed Gaussian Ansatz is a clear improvement and yields reasonable results at the whole range of energies, where the harmonic approximations begin to fail.

\subsection{Spontaneous Emission}

\begin{figure*}
  \centering
    \includegraphics[width=0.9\textwidth]{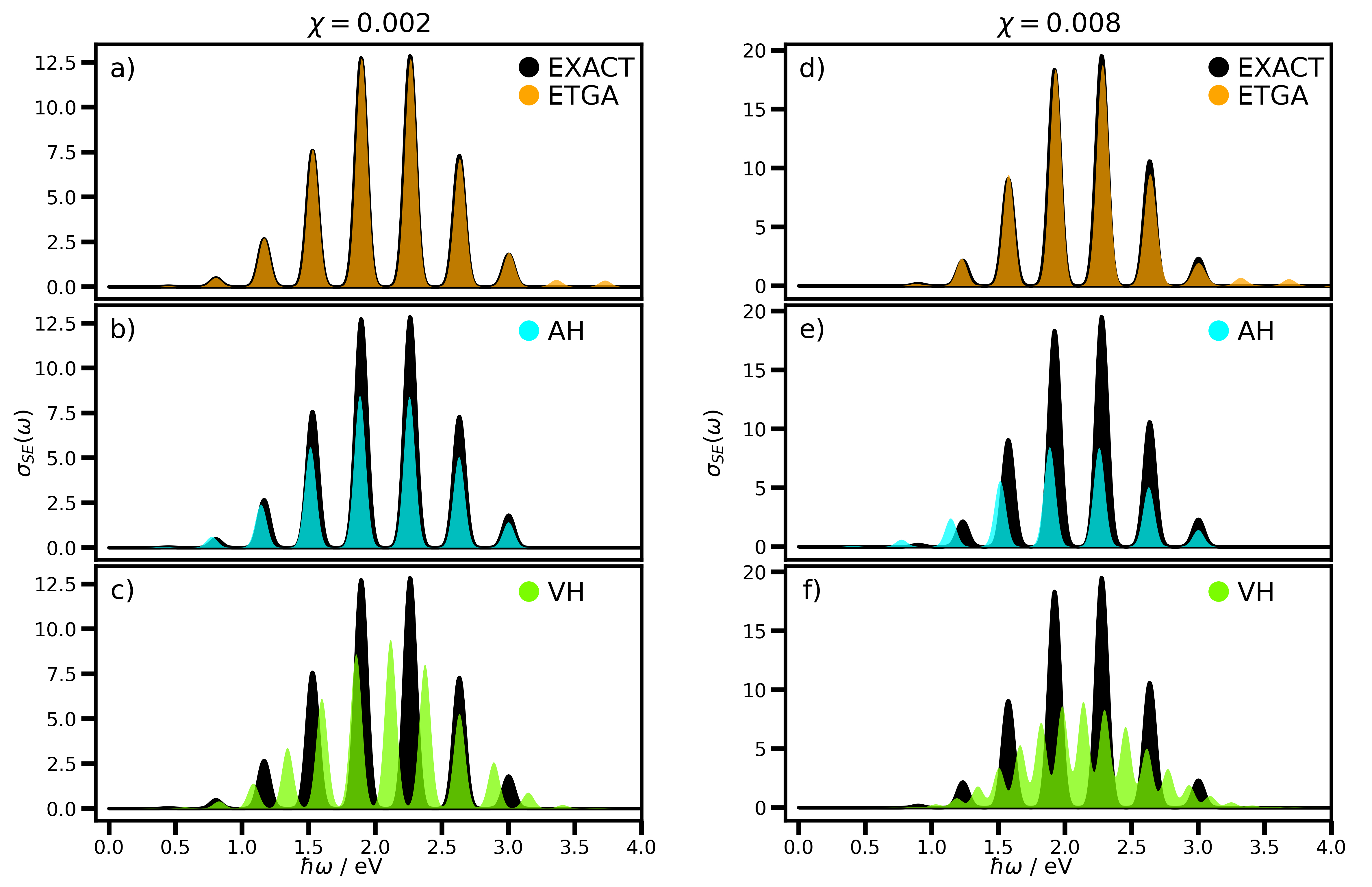}
\caption{Spontaneous emission spectrum based on a Morse potential with anharmonicity of 0.002 (left column, a--c) and anharmonicity 0.008 (right column, d--f). The reference spectrum in black is based on the Morse potential without approximations. The top panels (a,d) show the ETGA spectrum in orange, the adiabatic harmonic model (cyan) is in the middle (b,e) and the vertical harmonic model (green) at the bottom (c,f). All spectra have been broadened using a Gaussian line shape function with a HWHM of 0.05 eV. The corresponding emission rates are obtained by integration of the spectra.\label{fig:morse_emission_002_008}}
\end{figure*}

The numerical values of the transition dipole moment \(\mu_0\) and its derivative \(\mu'\) are chosen for all calculations such that \({\frac{\mu_0 }{(6\pi^2\hbar \varepsilon_0 c^3)^{1/2}}=10 \ \mathrm{a.u.} }\) and \({\frac{\mu' }{(6\pi^2\hbar \varepsilon_0 c^3)^{1/2}} =-0.5 \ \mathrm{a.u.}}\) These values yield emission rates of similar magnitude to the internal conversion rates, which is necessary in order to obtain reasonable quantum yields. The emission spectra for the Morse potentials with low and high anharmonicity are given in Figures \ref{fig:morse_emission_002_008}a--c and \ref{fig:morse_emission_002_008}d--f, the intermediate case with anharmonicity \(\chi=0.004\) is not explicitly shown.

The 0-0 transition corresponds to the peak at 3.0 eV, while transitions to states with a higher vibrational quantum number are found to the left of it. The intensity of peaks corresponding to transitions to states with a small energy gap to the initial state, decreases and approaches zero as the gap closes due to the scaling of spontaneous emission with \(\omega^3\). This scaling means that the prediction of states with a noticeable gap is more relevant when it comes to the prediction of emission rates, unlike internal conversion rates, where an accurate description of states with a vanishing energy gap is essential. We can observe that the ETGA (Fig. \ref{fig:morse_emission_002_008}a,d) provides overall a good description of intensities and frequencies. The method however produces weak erroneous peaks at frequencies that are too high, wrongly implying the existence of states below the vibrational ground state of the final potential.

This problem is also observed in the vertical harmonic model (Fig. \ref{fig:morse_emission_002_008}c,f), but unlike the ETGA, it shows poor performance  predicting intensities and frequencies over the whole range of transition frequencies. The adiabatic harmonic model (Fig. \ref{fig:morse_emission_002_008}b,e) on the other hand gives reasonable results close to the the 0-0 transition and benefits from the suppression of peaks at small energy gaps, which hides the shortcomings of the model in describing highly excited vibrational states. It also lacks the erroneous peaks at high transition energies but the intensities and the peak spacing are not as good as in the case of the ETGA model. 

\subsection{Quantum Yields}

We can now proceed to the calculation of radiative quantum yields. Integration of the emission spectra yields the spontaneous emission rate, and the internal conversion rate is given by the value of the internal conversion spectrum evaluated at zero energy. The quantum yield is then obtained using Eq. (\ref{eq:quantum_yield}), giving rise to the results summarized in Table \ref{tab:all_anharm}.

\begin{table}
\centering
\caption{\label{tab:all_anharm}Internal conversion rate, spontaneous emission rate and radiative quantum yield for varying degrees of anharmonicity \(\chi\) of a Morse potential.}
\begin{ruledtabular}
\begin{tabular}{rcccc}  
$\chi=0.002$ &  EXACT & ETGA & AH & VH \\
$k_\text{IC}  / (\mathrm{E_h/\hbar})$ & \num{3.94e-01} & \num{4.24e-01} & \num{5.93e-01} & \num{7.50e-01}  \\
$k_\text{SE} / (\mathrm{E_h/\hbar})$ & \num{1.78e-01} & \num{1.81e-01} & \num{1.25e-01} & \num{1.81e-01} \\
$\Phi_\text{QY}\times 10^2$ & 31.1 & 29.9 & 17.4 & 19.5\Bstrut\\
\hline
$\chi=0.004$ &  EXACT & ETGA & AH & VH \Tstrut  \\
$k_\textrm{IC} / (\mathrm{E_h/\hbar})$ & \num{1.79e-02} & \num{4.40e-02} & \num{5.93e-01} & \num{4.31e-01} \\
$k_\textrm{SE} / (\mathrm{E_h/\hbar})$ & \num{2.04e-01} & \num{2.08e-01} & \num{1.25e-01} & \num{2.08e-01} \\
$\Phi_\textrm{QY}\times 10^2$ & 91.9 & 82.5 & 17.4 & 32.5\Bstrut\\
\hline
$\chi=0.008$ &  EXACT & ETGA & AH & VH \Tstrut \\
$k_\textrm{IC} / (\mathrm{E_h/\hbar})$ & \num{1.19e-02} & \num{3.06e-03} & \num{5.93e-01} & \num{5.23e-02}  \\
$k_\textrm{SE} / (\mathrm{E_h/\hbar})$ & \num{2.44e-01} & \num{2.48e-01} & \num{1.25e-01} & \num{2.48e-01} \\
$\Phi_\textrm{QY}\times 10^2$ & 95.3 & 98.8 & 17.4 & 82.6 \\
\end{tabular}
\end{ruledtabular}
\end{table}

The ETGA is an improvement over the global harmonic models. The emission rates are in excellent agreement with the exact values, just like the emission spectra in Fig. \ref{fig:morse_emission_002_008}a/d. The internal conversion rates are also in good agreement but the spectra (Fig. \ref{fig:ic_spectra_morse_med_high}) also show  the downside of using a time dependent ansatz. The ETGA does not guarantee positive valued spectra. This can be problematic for internal conversion rates, which are evaluated at a single frequency if we adhere strictly to energy conservation. The method does deliver the best agreement if we judge the spectrum as a whole but it can give nonphysical results at specific energies. The method also suffers from decreasing accuracy as the anharmonicity increases. It does yield the best results however and the overall agreement of both spectra --- emission and internal conversion--- allows for a higher degree of confidence than pure harmonic models, especially in the case of internal conversion where the anharmonicity of the potential needs to be taken into account.

The emission spectra based on the adiabatic harmonic approximation show notable deviations in the peak intensities as the anharmonicity increases, but the predicted frequencies are reasonable. The decrease of the dissociation energy that is used to raise the anharmonicity has no effect on the adiabatic approximation due to the constant second derivative at the equilibrium position. The AH model yields in all cases the same results, independent of the increasing anharmonicity. It is adequate for transitions to the first few vibrational states of the final potential, but the energy of higher lying states is overestimated. This leads to large deviations in the case of the internal conversion spectra, where the energy conserving transition ends in an highly excited vibrational state. 

A quick glance at the quantum yields suggests that the VH model is superior to the AH model. The integrated areas of the emission spectra are close to the exact ones, due to the VH method's ability to capture the correct short-time dynamics, which determines the overall envelope and area. Due to this fact, the integrated emission rates are virtually identical to the results of the ETGA method, with the benefit of a much lower computational cost. But the internal conversion rate poses some difficulty. It is clear that the individual vibrational states of the final potential are incorrect in the vertical model, which can cause deviations regarding the internal conversion rate. This is due to the fact that the IC rate is not obtained by integration but determined by the detailed shape of the spectrum at the point of energy conservation. The vibrational frequency for the vertical harmonic approximation is derived at a position where the Morse potential shows more gentle incline. As the anharmonicity increases, the frequencies decrease further and the energy gap between states gets underestimated, leading to spectra with closely spaced peaks which can lead to a coincidental intersection of the vertical harmonic and exact spectra at the relevant value. It is not possible to predict a priori whether the VH model matches the true value or not for the internal conversion rate as long as the vibronic states remain separated. The AH model on the other hand is correct for the first few vibrational states but most likely wrong if there is a notable energy gap between initial and final state.

\section{Symmetric Double-Well Potential}

\subsection{Internal Conversion and Emission Spectra}

The ETGA was already applied to simulate wave packet dynamics on double-well potential energy surfaces of Pnictogen hydride cations to obtain vibronically resolved photoelectron spectra\cite{Begusic2022}. It was shown that the quality of the results varies, depending on the width and initial position of the wave packet with respect to the double-well. The method is  used here to predict internal conversion rates, to test its ability and reliability to handle floppy molecules when it comes to the estimation of quantum yields.

The final potential is assumed to be  a symmetric double-well defined by
\begin{equation}
    \begin{aligned}
        V_f(q) &= \frac{b}{q^4_{f,0}}(q-q_{f,0})^2(q+q_{f,0})^2 + V_{f,0}  \\
    \end{aligned}
\end{equation}

\begin{figure}
  \centering
  \includegraphics[width=.50\textwidth, trim=0 0.25cm 0 0.25cm, clip]{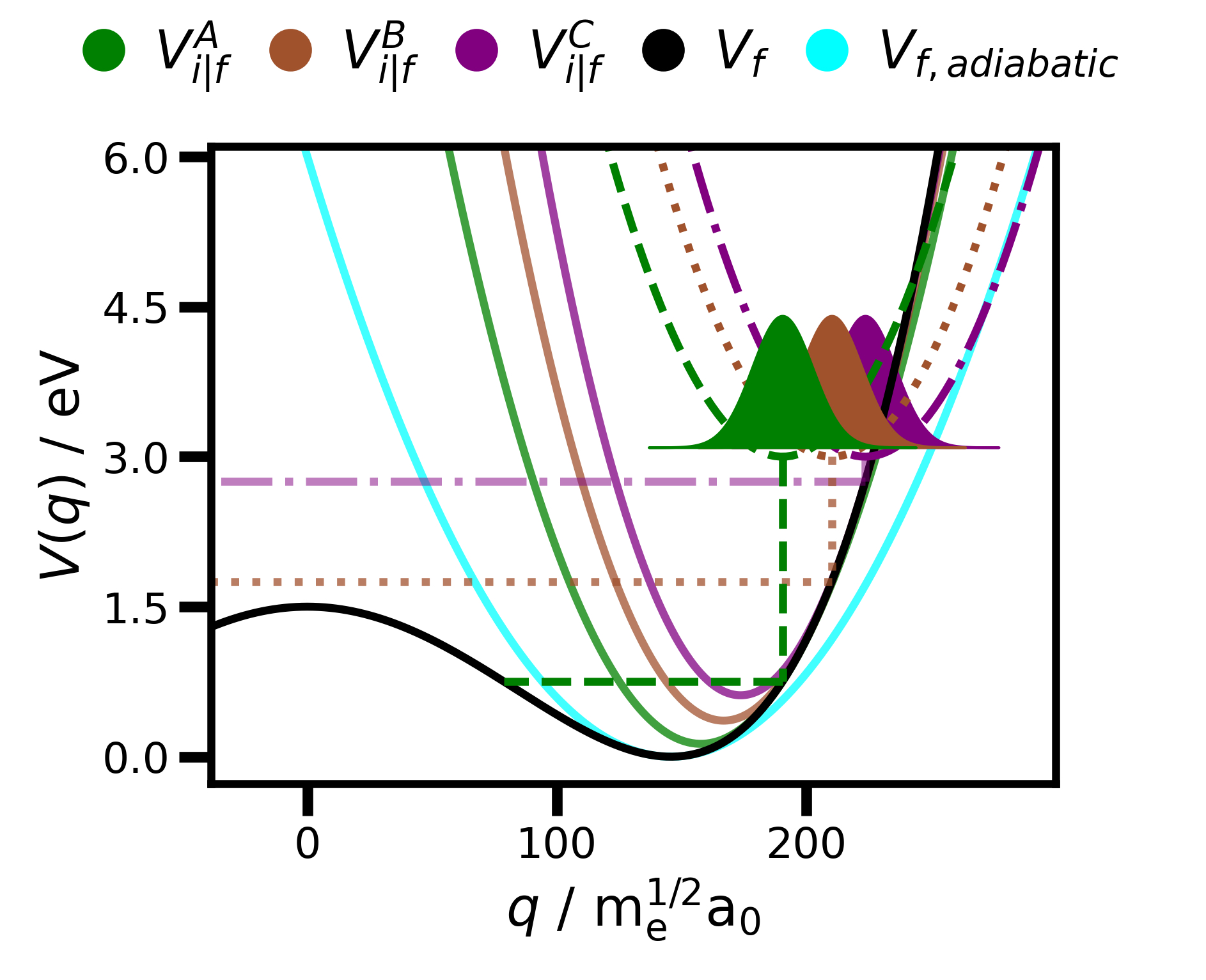}
\vspace{-20pt}
  \caption{Symmetric double well potential and three shifted harmonic potentials that are used for the simulation of internal conversion and spontaneous emission. The horizontal lines indicate the initial position and turning point of classical trajectorie that start at the equilibrium positions of the  initial harmonic potentials \(V_i^A, V_i^B, \textrm{ and } V_i^C\). The adiabatic harmonic approximation \(V_{f,adiabatic}\) of the double well potential \(V_f \) is constructed by harmonic expansion around the closest minimum of the double well with respect to the initial potential minimum. The potentials for the vertical harmonic model are also shown and indicated as \(V_f^A, V_f^B\) and \(V_f^C\).}
  \label{fig:double_well_potential_with_initial_potentials}
\end{figure}

The minima of the double well are at \({  q_{f,0}= \pm 145.747 \ \mathrm{ m_e^{1/2}a_0} }\) and the energetic barrier between them is given by \(b=1.5 \textrm{ eV}\). This choice of parameters leads to an angular frequency \(\omega_f=1000 \textrm{ cm}^{-1}\) in a harmonic approximation at \(q_{f,0}\) for the adiabatic harmonic model. While the final potential remains the same, shifted harmonic potentials with angular frequency \(\omega_i = 1500 \textrm{ cm}^{-1} \) and \(V_0=3.0 \textrm{ eV} \) are used to define the initial wave packet. The setup consists of three cases, with \(q_{0,A} = 190.428  \ \mathrm{ m_e^{1/2}a_0},\)  \(q_{0,B} = 210.206  \ \mathrm{ m_e^{1/2}a_0},\)  and  \(q_{0,C} = 223.616  \ \mathrm{ m_e^{1/2}a_0}.\) The initial positions were chosen based on the associated classical trajectories in the final double well potential. The classical energy for a trajectory starting at \(q_{0,A}\) is \(0.75 \textrm{ eV}\), half the value of the barrier, leading to a wave packet trapped to one side of the well. The energy at \(q_{0,B}\) is slightly above the barrier with \(1.75 \textrm{ eV}\) and the last case with \(q_{0,C}\) as initial position  corresponds to an energy of \(2.75 \textrm{ eV}\) considerably above the barrier.

\begin{figure*}
  \centering
  \includegraphics[width=\textwidth]{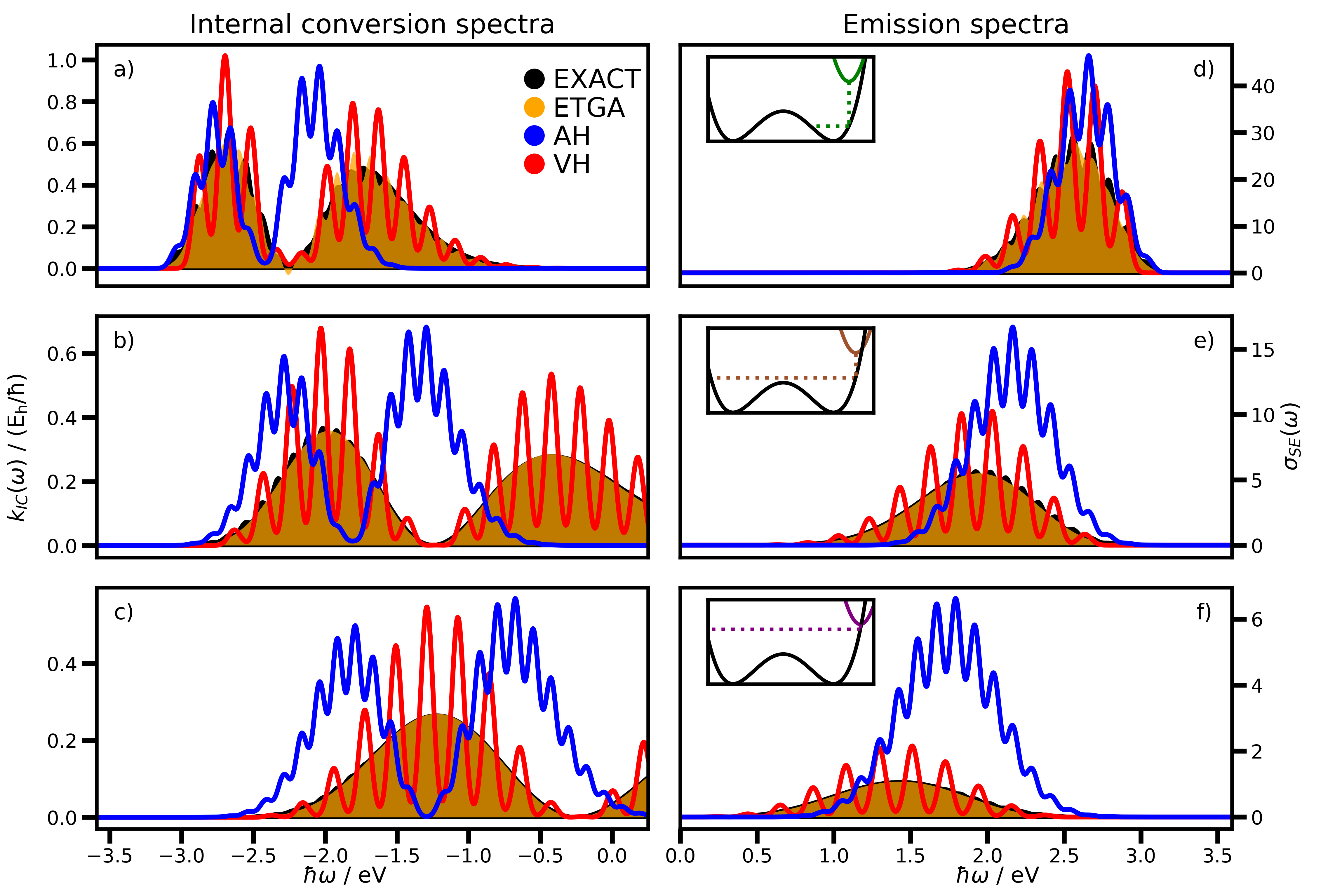}
  \caption{Internal conversion (left column, a--c)  and spontaneous emission (right column, d--f)  spectra of a double well potential (Fig. \ref{fig:double_well_potential_with_initial_potentials}). The position of the initial potential was chosen based on the potential energy of the classical trajectories in the double well. Position \(q_{0,A}\)  (a,d) leads to dynamics trapped to the initial side of the well, \(q_{0,B}\) (b,e) and \(q_{0,C}\) (c,f) corresponds to a trajectories with enough potential energy to cross the barrier as shown in the insets in the emission spectra. The reference spectra based on the exact potential are shown in black (filled), the ETGA results are given in orange (filled), the adiabatic harmonic model was used to obtain the blue-colored (line) spectra and the vertical harmonic model results are given in red. The internal conversion rates are given by the values of the IC spectra at \(\hbar\omega=0 \textrm{ eV}\), while the radiative rates are obtained by integration of the emission spectra. All spectra were broadened with a Gaussian line shape function with a HWHM of 0.05 eV.  \label{fig:double_well_spectra_005}}
\end{figure*}

\begin{figure*}
  \centering
  \includegraphics[width=\textwidth]{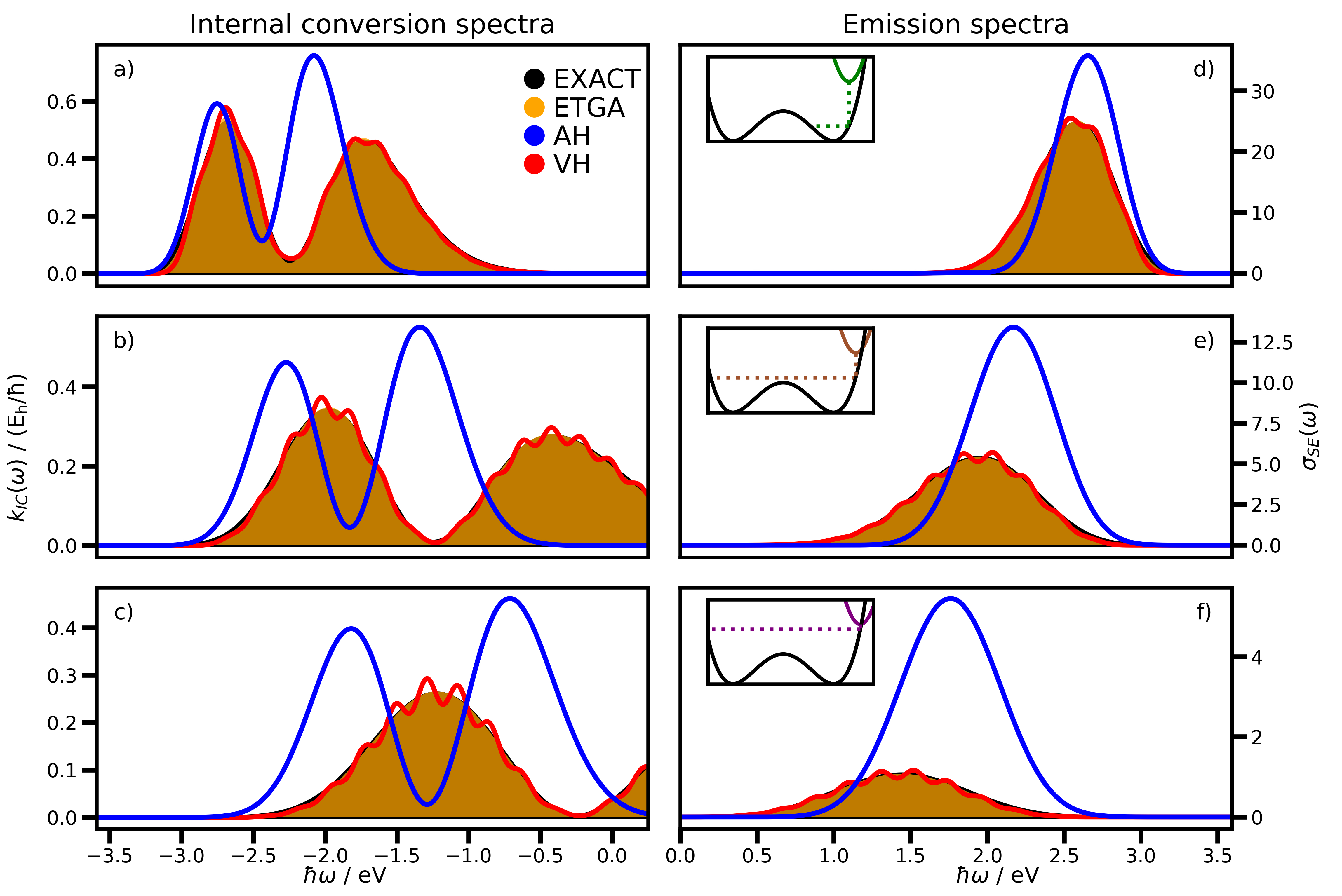}
  \caption{Internal conversion (left column, a--c)  and spontaneous emission (right column, d--f)  spectra of a double well potential (Fig. \ref{fig:double_well_potential_with_initial_potentials}). The position of the initial potential was chosen based on the potential energy of the classical trajectories in the double well. Position \(q_{0,A}\)  (a,d) leads to dynamics trapped to the initial side of the well, \(q_{0,B}\) (b,e) and \(q_{0,C}\) (c,f) corresponds to a trajectories with enough potential energy to cross the barrier as shown in the insets in the emission spectra. The reference spectra based on the exact potential are shown in black (filled), the ETGA results are given in orange (filled), the adiabatic harmonic model was used to obtain the blue-colored (line) spectra and the vertical harmonic model results are given in red. The internal conversion rates are given by the values of the IC spectra at \(\hbar\omega=0 \textrm{ eV}\), while the radiative rates are obtained by integration of the emission spectra. All spectra were broadened with a Gaussian line shape function with a HWHM of 0.1 eV.  \label{fig:double_well_spectra_01}}
\end{figure*}

\begin{figure*}
  \centering
  \includegraphics[width=0.95\linewidth]{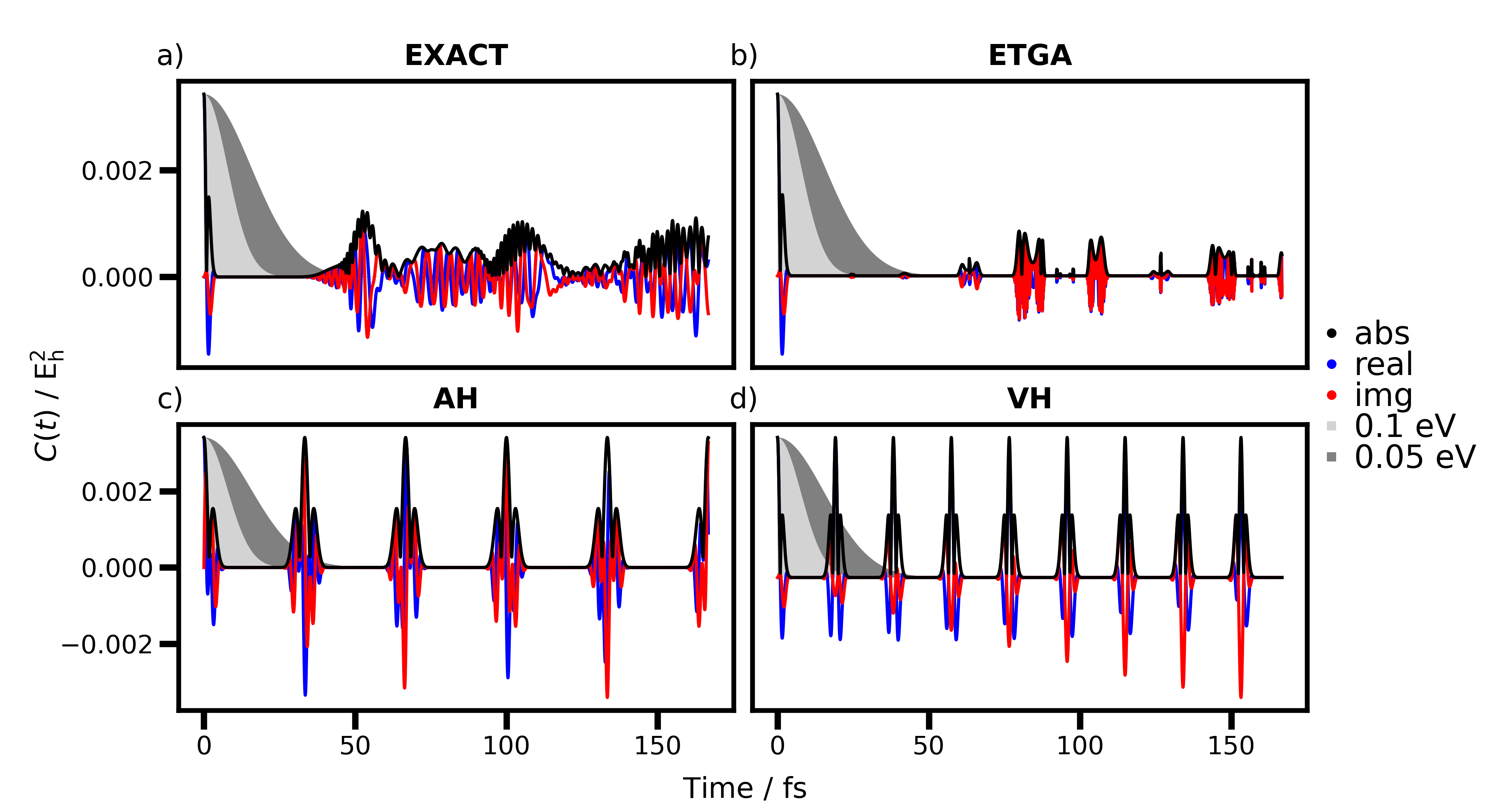}
  \caption{Auto correlation function \(C(t) =  \langle \phi_{i0} | U^\dagger_i T^\dagger_{fi}U_f T_{fi}|\phi_{i0}\rangle\) obtained by (a) numerical propagation using a split-operator scheme, (b) the extended thawed Gaussian Ansatz , (c) the adiabatic harmonic approximation and (d) the vertical harmonic approximation for the transition from \(V_i^C\) to the double well potential as shown in Figure \ref{fig:double_well_potential_with_initial_potentials}. The time-domain Gaussian envelopes, corresponding to a HWHM of 0.1 eV and 0.05 eV in the frequency domain, are shown in light gray and gray as background.}
  \label{fig:cor_double_well_C}
\end{figure*}

A second order Taylor expansion of the final potential for each initial position yields the harmonic potentials used in the vertical harmonic model. The frequencies of these harmonic oscillators are \(\omega_A \approx 1435 \textrm{ cm}^{-1}, \omega_B \approx 1619 \textrm{ cm}^{-1}\) and \(\omega_C \approx 1741\textrm{ cm}^{-1}.\)
Figure \ref{fig:double_well_potential_with_initial_potentials} shows the potentials, the initial state wave function and the classical energy. The wave packets are again propagated using the exact potential in a split-operator scheme, the ETGA,  the VH and the AH model. The transition dipole moment coupling parameters are \({\frac{\mu_0 }{(6\pi^2\hbar \varepsilon_0 c^3)^{1/2}}=10 \textrm{ a.u.}}  \) and  \({\frac{\mu' }{(6\pi^2\hbar \varepsilon_0 c^3)^{1/2}}=-0.5 \textrm{ a.u.} }\) The kinetic coupling for internal conversion is set to \(\tau_{fi} = 1.0 \ \mathrm{(m_e^{1/2}a_0)^{-1}}.\) The wave packets were in all cases propagated for \(T=10\pi/\omega_f\approx 166 \textrm{ femtoseconds}\) to obtain the correlation functions. 

In this case we also explore how the radiative and nonradiative rate vary as a function of the width of a Gaussian envelope, which determines the damping of the correlation function from a time-domain point of view. With this we can assess how the method performs as we look at the statistical limit for relaxation\cite{Siebrand1967,Bixon1968,Freed1970,Nitzan1973,Nitzan1973a,Valiev2020}, the case where the lifetime of the final states is short. This corresponds to the assumption that the relaxation of the final states is faster than the recurrence time of the wave packet which is assumed to be the case in large poly-atomic molecules or systems in contact with an environment. In either case one can assume that there is a high density of states that couple to the final vibronic states of the molecule, leading to a fast vibrational relaxation that limits the lifetime. This is in contrast to the former example of the Morse potential where we included recurrences of the wave packet and assumed a long lifetime of the final vibrational states. For our one-dimensional model we assume that the broadening is due to an environment and use thus a Gaussian line shape.
We begin with a small spectral width of 0.05 eV for the Gaussian HWHM to see the vibronic structure due to individual states. The results for internal conversion and emission are shown shown in Figure \ref{fig:double_well_spectra_005} side by side. 

The internal conversion spectra given in Figure \ref{fig:double_well_spectra_005}a are based on the trapped wave packet. The internal conversion rate is the value of the spectrum at energy conservation, i.e. an energy difference of final and initial state of zero. The rate value is negligible in this case but the internal conversion spectrum shows that the ETGA (orange filled) provides a better approximation to the exact reference spectrum (black filled) as the energy gap closes while the adiabatic harmonic approximation (blue line) is better suited to model the transition to the vibrational ground state of the final potential. The vibronic structure of the vertical harmonic model is quite pronounced but does not match the real states, however the general shape is in fair agreement and does not deviate as strongly as the adiabatic model as the energy increases.
The corresponding emission spectra are given in  Figure \ref{fig:double_well_spectra_005}d. The adiabatic harmonic approximation spectrum seems to be superior to the ETGA in the prediction of the 0-0 transition at 3.031 eV but starts to overestimate the intensities for transitions to higher vibrational states compared to the exact reference spectrum. The ETGA spectrum however yields decent results as the energy gap between final and initial states grows smaller. 

Figures \ref{fig:double_well_spectra_005}e,f depict the emission spectra of wave packets that can cross the barrier. The ETGA emission spectra are lacking any separation of vibronic states, but the general shape and intensities are close to the exact result in both cases. The explanation for this is the ETGA's ability to capture the initial, short-time wave packet dynamics. This yields good results for the central frequencies and envelopes of the observed peak patterns, but a prediction of the fine splitting requires a good description for a longer duration, when the thawed Gaussian approximation for the wave packet would fall off in quality with time. The adiabatic approximation largely overestimates the intensities and the constant peak spacing of the harmonic model leads to growing errors in the frequency for transitions to higher lying vibrational states of the double well potential. This is especially problematic for the internal conversion rate, seen in Figures \ref{fig:double_well_spectra_005}b,c. Here, the general shape of the adiabatic model is a poor fit to the reference. The curve of the adiabatic harmonic model crosses the reference spectrum at some points, but the agreement of the values is clearly accidental and not systematic nor reliable, barring the 0-0 transition. The vertical harmonic model is also superior to the adiabatic approximation although the pronounced vibronic structure is clearly undesirable in this case, as it predicts noticeable fluctuations in the internal conversion rate with small energy shifts.

We proceed by increasing the HWHM of the spectral envelope from the former 0.05 eV to 0.1 eV, corresponding to a damping of the correlation function such that recurrences are almost suppressed. In this case the spectrum is  determined by the initial decay of the correlation function, given by the time the wave packet needs to leave the region of overlap initially. The results are gathered in Figure \ref{fig:double_well_spectra_01}. Individual peaks can no longer be distinguished as the vibronic structure is smoothed out. The splitting due to recurrences in the correlation function is gone and the form of the spectrum is dominated by the envelope due to the fast initial decay time. The ETGA and vertical harmonic model yield almost the same results and it becomes clear that they will converge for even higher values of the HWHM. The time-dependent potential used in the ETGA is approximately constant on this short timescale and at the start of the dynamics it is identical to the potential used in the vertical harmonic model. There clearly is a time-scale on which both models are barely distinguishable and  the models will converge to the same results if the initial decay occurs on this time-scale. The adiabatic harmonic model is inferior since it is generally not the best harmonic approximation to the true potential around the point where the initial dynamics takes place; barring the case where the initial and final equilibrium position are identical. The time scale is easily identified in Figure \ref{fig:cor_double_well_C}, showing the internal conversion correlation function for initial potential C. There are no simple recurrences in the exact correlation function (Fig.\ref{fig:cor_double_well_C}a) as the wave packet dynamics is quite complicated and far from the harmonic case. The time-domain correlation function (Fig.\ref{fig:cor_double_well_C}b) also shows that the ETGA method is not able to capture the long-term dynamics correctly in this case, meaning that it could not resolve the vibronic structure even if we had chosen a very small spectral line width. The time for the initial decay is clearly overestimated in the AH model (Fig. \ref{fig:cor_double_well_C}c), explaining its failure to predict the envelopes of the spectra. The VH model (Fig. \ref{fig:cor_double_well_C}d) on the other hand shows good agreement on the shortest time scale, to which the dynamics is restricted when damped by the Gaussian corresponding to a spectral width of 0.1 eV. The HWHM value of 0.05 leads to a damping that includes recurrences of the correlation function in the VH model, explaining the distinct splitting of the peaks.

\begin{figure}
  \centering
  \includegraphics[width=.40\textwidth, trim=0 0.0cm 0 0.0cm, clip]{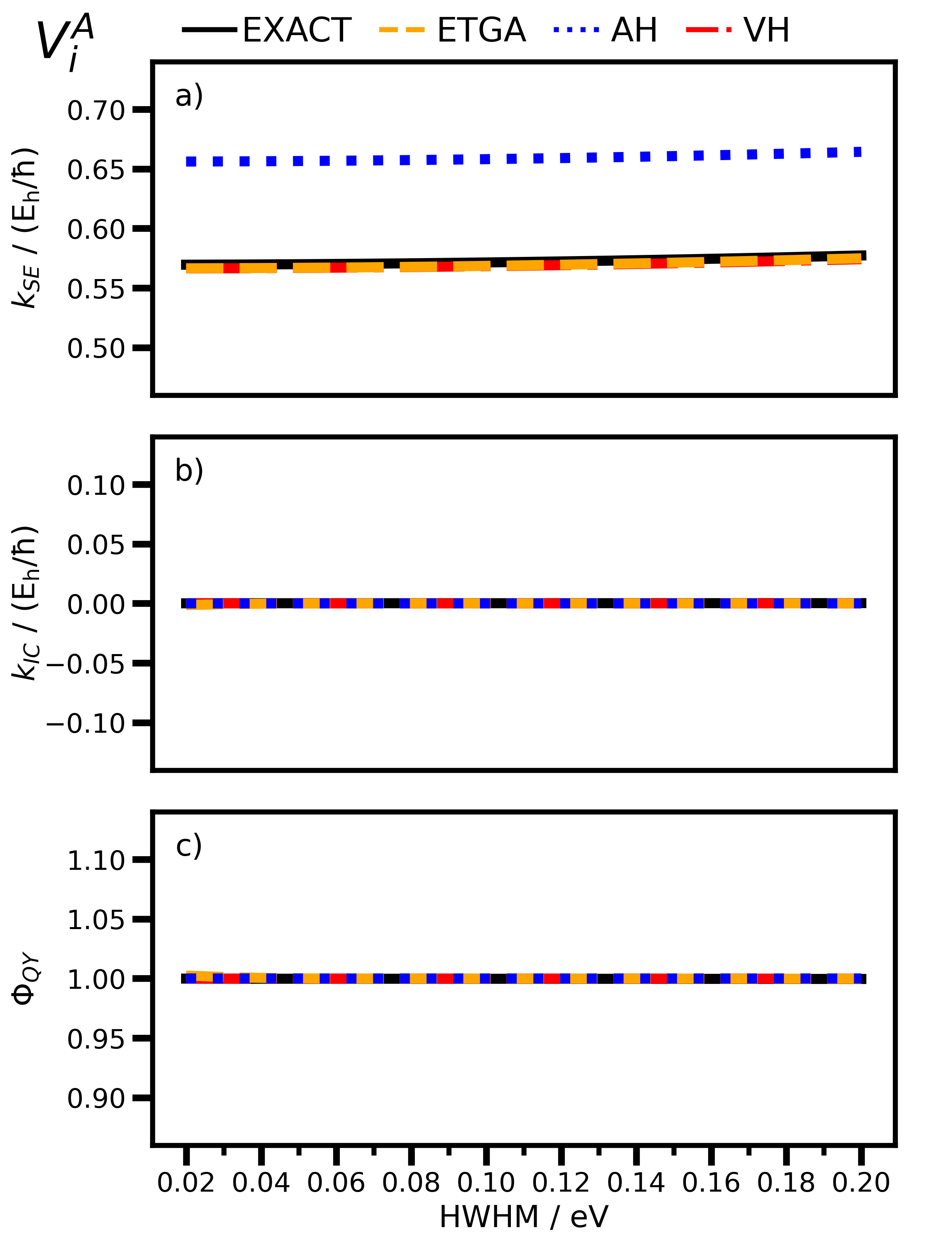}
\vspace{-10pt}
    \caption{Spontaneous emission rate (a), internal conversion rate (b) and quantum yield (c) as functions of the HWHM of the applied Gaussian damping function, for the transitions from initial potential \(V_i^A\) to a double well potential, calculated using a exact numerical propagation scheme, the ETGA, AH and VH approximation. \label{fig:qy_vs_hwhm_a}}
\end{figure}

\subsection{Quantum Yields, Internal Conversion and Emission Rates}

\begin{figure}
  \centering
  \includegraphics[width=.45\textwidth, trim=0 0.0cm 0 0.0cm, clip]{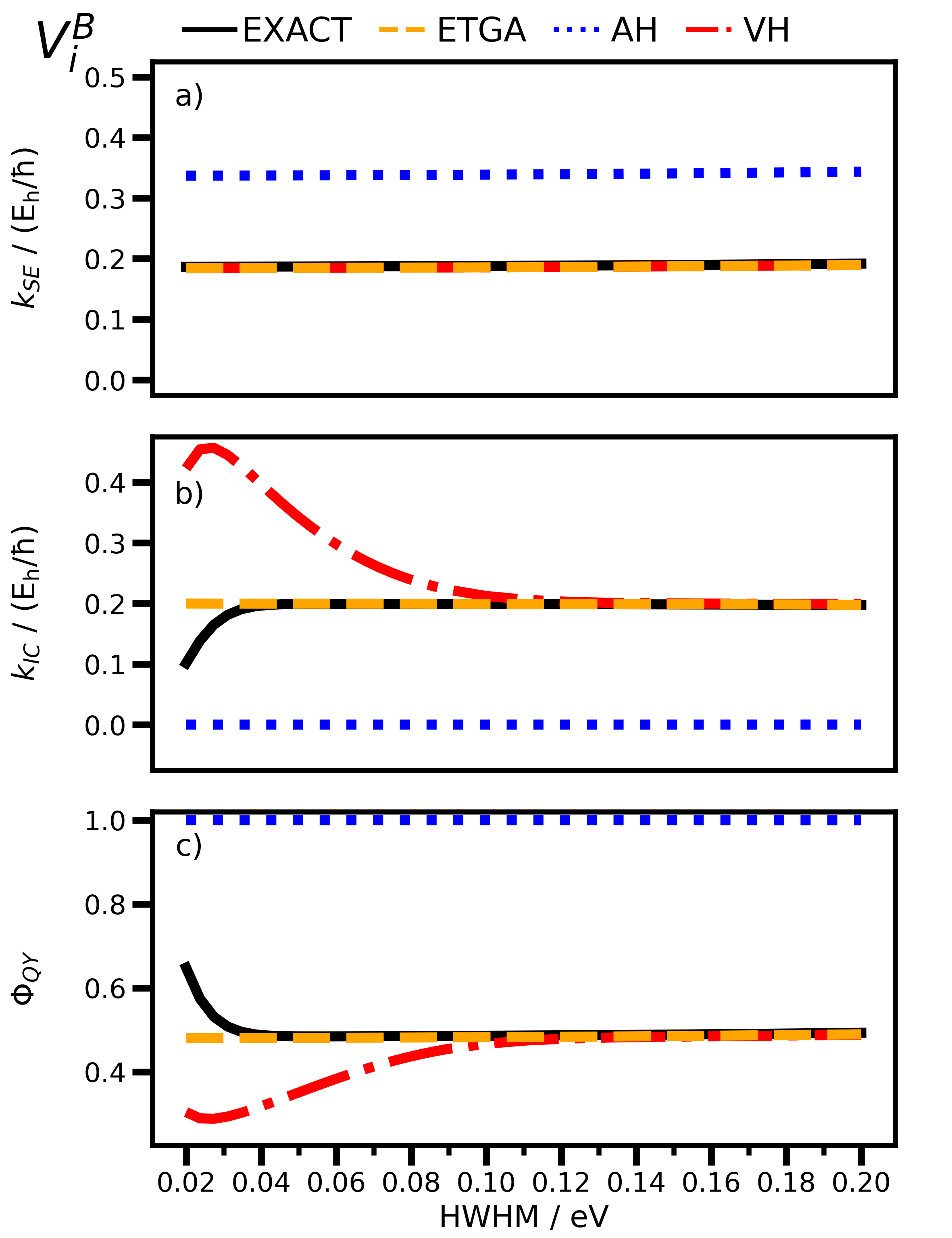}
\vspace{-10pt}
    \caption{Spontaneous emission rate (a), internal conversion rate (b) and quantum yield (c) as functions of the HWHM of the applied Gaussian damping function, for the transitions from initial potential \(V_i^B\) to a double well potential, calculated using a exact numerical propagation scheme, the ETGA, AH and VH approximation. \label{fig:qy_vs_hwhm_b}}
\end{figure}

\begin{figure}
  \centering
  \includegraphics[width=.45\textwidth, trim=0 0.0cm 0 0.0cm, clip]{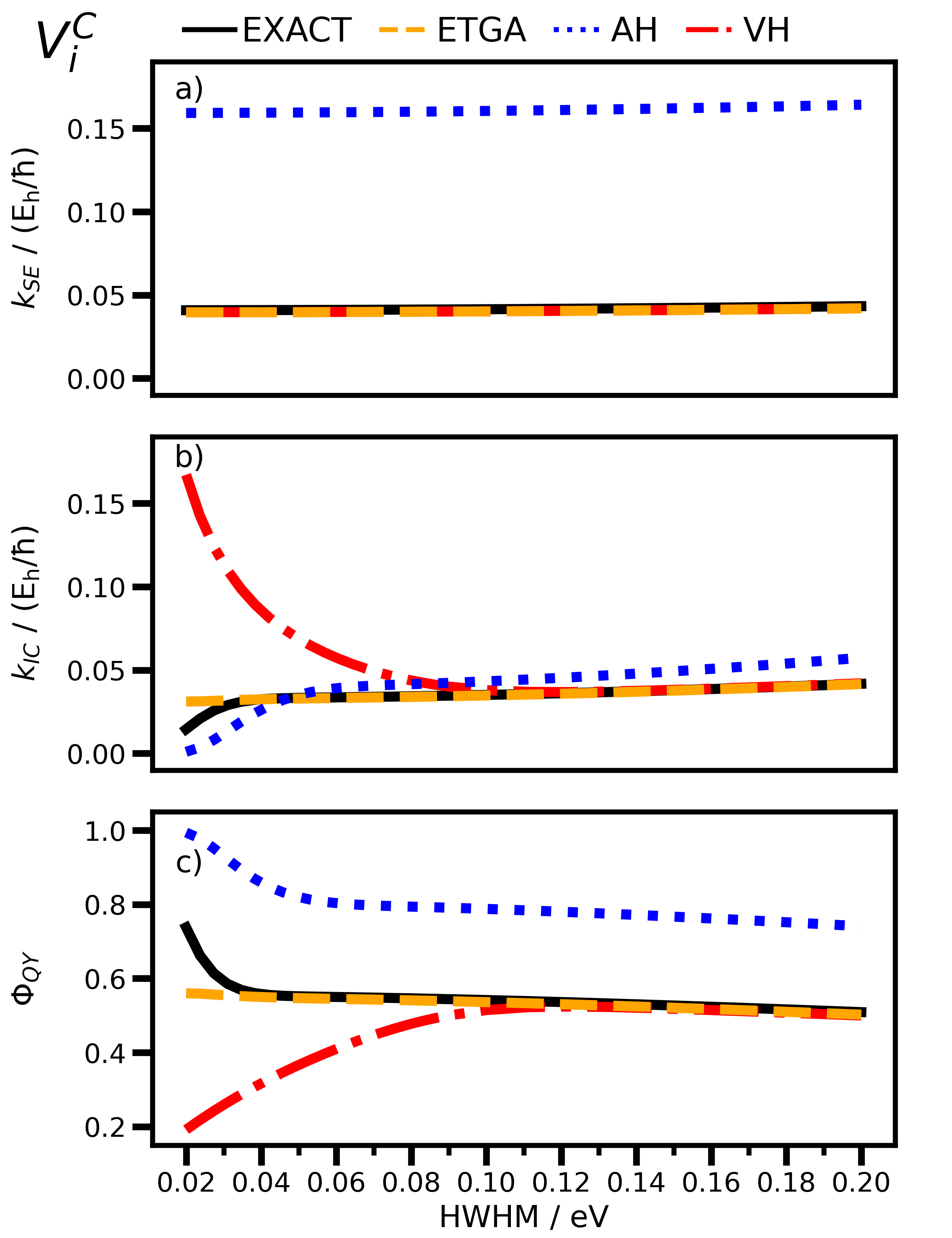}
\vspace{-10pt}
    \caption{Spontaneous emission rate (a), internal conversion rate (b) and quantum yield (c) as functions of the HWHM of the applied Gaussian damping function, for the transitions from initial potential \(V_i^C\) to a double well potential, calculated using a exact numerical propagation scheme, the ETGA, AH and VH approximation. \label{fig:qy_vs_hwhm_c}}
\end{figure}

The emission and internal conversion rates for the transitions from initial potential \(V_i^A\) are quantified in Figure \ref{fig:qy_vs_hwhm_a} and used to calculate the fluorescence quantum yield according to Eq.  (\ref{eq:quantum_yield}). The values are plotted as functions of the of the frequency-domain Gaussian HWHM. The emission rate is basically constant as it is obtained by integration of the whole emission spectrum, while the internal conversion rate remains fixed at zero since there is no peak near the point of energy conservation that could cause a rise when broadened. The one thing that stands out is the result of the AH model for the emission rate, which is clearly overestimated. The ETGA and the VH model yield the same results for the total emission rate after integration of the spectrum (Fig \ref{fig:double_well_spectra_005}d, \ref{fig:double_well_spectra_01}d).

The results starting from potential \(V_i^B\) are given in Figure \ref{fig:qy_vs_hwhm_b}. The emission rate (Fig. \ref{fig:qy_vs_hwhm_b}a) is again invariant to the broadening and the VH and ETGA model yield equivalent results once the spectra are integrated, which are in good agreement with the exact value. But the internal conversion rate (Fig. \ref{fig:qy_vs_hwhm_b}b), which is not obtained by integration but by a single value, shows that there is a regime in which the ETGA method and the vertical harmonic model behave differently. At the lowest values of the HWHM around 0.02 to 0.04 eV, separated individual vibronic states would begin to show. The ETGA method is not capable to model the necessary long time dynamics accurately enough to obtain such a fine resolution in the energy domain and deviates from the exact results in this regime. The vertical harmonic model also predicts a wrong spacing of vibronic states and deviates as well. But the exact result quickly approaches a constant value once the broadening leads to a continuous, overlapping spectrum. The vertical harmonic model also predicts this value in the short-time limit that corresponds to a dampening of the correlation function such that only the initial decay matters, beginning approximately at a broadening of 0.1 eV. The adiabatic harmonic model is wrong for all values of broadening and predicts a quantum yield of 1.0 for all values due to an erroneous internal conversion rate of zero.

A similar  behaviour is observed for the last case with the transition originating from initial potential \(V_i^C\). The emission rate (Fig. \ref{fig:qy_vs_hwhm_c}a) is again overestimated by the AH model but, unexpectedly, the result for the internal conversion rate (Fig. \ref{fig:qy_vs_hwhm_c}b) is now similar to the exact result for all values of the HWHM. However, a look at the internal conversion spectra given in Figures \ref{fig:double_well_spectra_005}/\ref{fig:double_well_spectra_01}c shows that this is only the case due to a crossing of the adiabatic internal conversion spectrum with the exact spectrum at the point of energy conservation at 0.0 eV. The internal conversion spectra indicate that this is a lucky coincidence and the rate value would clearly deviate if the energy shift of the potentials was slightly different, moving the point of energy conservation away from this intersection. The trends in the ETGA and VH model are again the same as in the case before. The ETGA method agrees with the exact results after a small broadening and the VH method converges to the same values as the exact and ETGA model once the broadening corresponds to such a  damping in the time domain that only the initial decay of the correlation function matters.

\section{Conclusion}

In this work, the semi-classical Extended Thawed Gaussian Approximation was for the first time applied to predict internal conversion rates. We have illustrated the method's validity and compared its performance with the adiabatic and vertical harmonic models by simulations of internal conversion and emission spectra in increasingly anharmonic potentials, starting with three Morse potentials followed by a double-well potential.

The investigation of the Morse potentials showed the following: The adiabatic harmonic model is capable of treating transitions to the lowest vibrational state, which is sufficient for emission spectra since these transitions usually constitute the largest contribution, also enhanced by the weighting with \(\omega^3\) which intensifies peaks at the highest transition energy. But the model is insufficient when it comes to transitions to higher lying vibrational states, which is particularly important for internal conversion.

The ETGA method proves superior and is able to capture effects on the transition amplitudes and frequencies caused by the anharmonicity of the underlying potential. Despite being a clear improvement with respect to harmonic models, it also has difficulties with increasing anharmonicity. Negative peaks and signals at the wrong frequency grow in intensity. The adiabatic model however remains completely oblivious to the growing anharmonicity in the case of the Morse potential, as the only relevant parameter, the second derivative of the potential at the equilibrium, remains constant. The AH model thus predicts the same results for all investigated values of anharmonicities, clearly a big disadvantage.

The vertical harmonic model is not invariant to the changes of anharmonicity and it yields in all cases very good values for the emission rate, as the latter is rather insensitive to the vibrational structure of the integrated emission spectrum as long as the outline and position of the exact spectrum are matched. The vertical harmonic model is capable to do so by capturing the initial decay of the correlation function accurately, but the vibrational structure and the splitting of individual peaks depend on the recurrence time after the initial decay of the correlation function, which cannot be obtained within the VH model. This is a problem if vibrational relaxation or dephasing due to an environment are not sufficiently fast to suppress the vibrational structure, equivalent to a fast damping of the correlation function in the time domain.

The limit of strong damping and a fast decay of the correlation function was considered for the double well potential by applying a Gaussian damping functions with varying width. Differences between the ETGA and the VH results vanish in all cases considered once the damping limits the dynamics to the initial decay, which corresponds roughly to a spectral broadening larger than 0.1 eV in this model. Both methods agree with the exact results in this limit and there is no benefit in using the ETGA over the VH model in this regime. The ETGA method however distinguishes itself from the VH model if the damping is reduced. The vertical harmonic model begins to diverge from the exact results once recurrences within the correlation function are included in the calculation. The same happens to ETGA model at even weaker damping as it also fails to reproduce the exact long term dynamics, which are required for a high spectral resolution. The adiabatic harmonic model yields the worst results. It generally overestimates the emission rate and values of the internal conversion rate are also unreliable if there is an appreciable energy gap between the potentials.

The AH model might be useful for internal conversion if the initial and final electronic states are almost degenerate in energy. But it cannot be used to model systems that emit at higher energies, entering the range of visible wavelengths, i.e. systems interesting for technical applications such as LEDs. The emission spectrum might be adequate but the internal conversion rate is most certainly wrong, unless the system is perfectly harmonic, in which case all models would predict the same. The vertical harmonic model appears to be a good method if only the initial decay of the correlation function is relevant. This should be the case if vibrational relaxation of the high-lying vibronic states of the final potential is fast, which should be a fair assumption for large polyatomic molecules or systems in contact with a bath that possesses a large density of states. The ETGA method is expected to be an improvement over the VH model if the lifetime of the final states is not too short. This is the case in optical transitions where the bright states are typically low lying vibronic states that have a long lifetime, which is accessible by measurements of the peak width in vibronically resolved spectra, explaining the methods benefits when it comes to absorption or emission spectra. But the lifetime of high-lying vibronic states, which are involved in internal conversion, are not readily available and one typically must make an assumption by choice of the width of the spectral lineshape function. The results showed that benefits of the ETGA are expected to play out in cases with an appreciable lifetime.

A robust and universally applicable method for the prediction of internal conversion rates and ultimately quantum yields is still amiss, but the ETGA is a useful extension to the commonly used harmonic models. Particularly for internal conversion rates in systems  where  transitions to higher vibrational states are relevant, the predictions are reliable if the anharmonicity is not too high. Even floppy systems can be treated. The vertical harmonic and the ETGA model are expected to yield the same results if the lifetime of vibrational states is very short such that only the initial decay of the correlation function matters. The ETGA is expected to provide better results if the lifetime is long enough such that recurrences need to be included.

The outlook for applications of the method to real molecules is promising. Although is has not been used yet to predict internal conversion rates, the method was already applied to simulate emission spectra of molecular systems.\cite{Prlj2020,Begusic2022} This showed the feasibility of the method for internal conversion rate calculations as the computational cost is almost identical to the cost of emission spectra simulations. The only difference is the need for the first order nonadiabatic coupling element instead of the transition dipole moment. But the same trajectory and the same Hessians can be used in the calculation, making up the biggest computational effort. It should also be noted that the formalism is  also applicable to system in thermal equilibrium and not limited to simulations at absolute zero, meaning that temperature effects can also be included.\cite{Begusic2020} The viability of the method was proven in this work and the transfer and application to real molecules is within reach.

\section*{Conflicts of Interest}

There are no conflicts to declare.

\begin{acknowledgments}
We gratefully acknowledge financial support by the Deutsche Forschungsgemeinschaft via grant MI1236/6-1.
\end{acknowledgments}

\section*{Data Availability}
The data that support the findings of this study are available from the corresponding author upon reasonable request.

\section*{References}
\nocite{*}
\bibliography{aipsamp}

\end{document}